\documentclass[a4paper,11pt]{article}
\usepackage{jheppub} 
\usepackage[utf8]{inputenc}
\usepackage{physics}
\usepackage{slashed}
\usepackage{caption}
\usepackage{xcolor}
\usepackage{comment}
\usepackage{multirow}
\usepackage{graphics}
\usepackage{float}
\usepackage{cancel}
\usepackage{soul}
\usepackage{array}
\usepackage{mathtools}  
\usepackage{amsfonts}

\title{\boldmath\boldmath A Foray on SCFT$_3$ via Super Spinor-Helicity and Grassmann Twistor Variables}

\author{Sachin Jain, Dhruva K.S, Deep Mazumdar and Shivang Yadav}

\affiliation{Indian Institute of Science Education and Research,\\ Dr Homi Bhabha Road, Pashan, Pune, India}

\emailAdd{sachin@iiserpune.ac.in}
\emailAdd{k.s.dhruva@students.iiserpune.ac.in}
\emailAdd{deepkamal.mazumdar@students.iiserpune.ac.in}
\emailAdd{shivang.yadav@students.iiserpune.ac.in}
\abstract{In this paper, we develop a momentum super space formalism for $\mathcal{N}=1,2$ superconformal field theories in three dimensions. First, we solve for super-correlators in the usual momentum superspace variables. However, we found that expressing quantities in super space spinor helicity variables greatly simplifies the analysis. Further, by performing a ``half" Fourier transform of the Grassmann coordinates which is analogous to the Twistor transform, an even more remarkable simplification occurs. Using these formalism, we first compute all three point correlation functions involving conserved super-currents with arbitrary spins in $\mathcal{N}=1,2$ theories. We discover interesting double copy relations in $\mathcal{N}=1$ super-correlators. Also, we discovered super double copy relations that take us from $\mathcal{N}=1$ to $\mathcal{N}=2$ super-correlators. We also comment on the connection of our results with the flat space super amplitudes in one higher dimension.}

\begin{document}
\maketitle
\section{Introduction}\label{intro}
Conformal field theory (CFT) forms a cornerstone of theoretical physics. It's applications range from string theory to condensed matter physics. A special class of CFTs are those that also possess supersymmetry, i.e super conformal field theories (SCFTs), see for instance \cite{Osborn:1998qu,Park:1999pd}. These theories are of utmost interest as they provide us a window into non-perturbative physics. It is quite advantageous to formulate such theories in momentum superspace. In non supersymmetric CFTs, momentum space has led to various interesting results such as making a connection to flat space scattering amplitudes in one higher dimension \cite{Maldacena:2011nz,Raju:2012zr}, discovering double copy structures at the level of correlation functions \cite{Farrow:2018yni,Lipstein:2019mpu,Jain:2021qcl} and even making a connection with early universe cosmology \cite{Maldacena:2011nz,McFadden:2011kk,Ghosh:2014kba,Arkani-Hamed:2015bza,Arkani-Hamed:2018kmz,Baumann:2022jpr} and AdS amplitude \cite{Raju:2010by,Raju:2012zs,Albayrak:2018tam,Albayrak:2019yve}. One natural direction is to explore supersymmetric extension, their connection to supersymmetric scattering amplitude, AdS superamplitude \cite{Bissi:2022wuh}, etc. While there has been a lot of development in the Fourier space approach to non-supersymmetric CFTs over the past decade or so \cite{Coriano:2013jba,Bzowski:2013sza,Bzowski:2015pba,Bzowski:2017poo,Bzowski:2018fql,Bautista:2019qxj,Jain:2020rmw,Jain:2020puw,Jain:2021wyn,Jain:2021qcl,Jain:2021vrv,Jain:2021gwa,Jain:2021whr,Isono:2019ihz,Gillioz:2019lgs,Baumann:2019oyu}, its supersymmetric counterparts have mostly been left untouched. The development of such a formalism would greatly aid for example the study of connection to flat space S-matrix. This development also have a application in Supersymmetric Chern-Simmons matter such as \cite{Inbasekar:2015tsa,Gur-Ari:2015pca,Aharony:2019mbc} and $\mathcal{N}=6$ ABJM theory\cite{Aharony:2008ug}.
However, such pursuits first require the development of momentum superspace. Most of the literature, however has been in the arena of position superspace \cite{Nizami:2013tpa,Buchbinder:2021gwu,Buchbinder:2021izb,Buchbinder:2021kjk,Buchbinder:2021qlb,Jain:2022izp,Buchbinder:2023fqv,Buchbinder:2023ndg}.
Thus, we begin the programme of the development of momentum superspace, starting with three-dimensional SCFTs which possess two or four supercharges.
\\\\
\textbf{Outline}:\\ In section \ref{sec:Neq1Formalism}, we introduce the momentum superspace formalism for $\mathcal{N}=1$ SCFTs. After working out a three-point super-correlator as an example, we show that rather than working in ordinary momentum superspace variables, working with super spinor helicity variables is quite advantageous. We then show that a drastic simplification occurs if we rather work in \textit{Grassmann twistor space}, forming the nexus of our formalism. In section \ref{sec:Neq1Corr}, we present our results for all two and three-point correlators with arbitrary (half) integer spin insertions in $\mathcal{N}=1$ SCFTs. We also discuss double copy relations as well as Ward-Takahashi identities. In section \ref{sec:Neq2}, we extend our formalism to theories with four supercharges as well as present a way to obtain $\mathcal{N}=2$ super-correlators via a super double copy construction of their $\mathcal{N}=1$ counterparts. Finally, in section \ref{sec:discussion}, we summarize our results and discuss several exciting future directions.\\ We also have several appendices that complement the main text. In appendix \ref{sec:notation}, we present our notations, conventions and some useful spinor-helicity identities. In appendix \ref{sec:algbera}, we present the $\mathcal{N}=1$ and $\mathcal{N}=2$ Lie superalgebras and the action of their generators on primary superfields in various variables. In appendix \ref{sec:KsusygivesK}, we prove that conformal invariance of the individual component correlators that appear in a super-correlator implies the conformal invariance of the super-correlator itself. Finally, we present a worked out example of a $\mathcal{N}$=1 three-point function in appendix \ref{sec:example}. 

\section{Setting the stage: $\mathcal{N}=1$ SCFT}\label{sec:Neq1Formalism}
 The coordinates of $\mathcal{N}=1$ superspace consist of a pair $(x^\mu, \theta^a)$ where $x^\mu, \mu=1,2,3$ are the usual position space coordinates and $\theta^a,a=1,2$ are the Grassmann coordinates of the superspace \cite{Park:1999cw, Nizami:2013tpa}. The generators of the $\mathcal{N}=1$ superconformal algebra that act on the superspace consist of the usual conformal generators $P_\mu, M_{\mu\nu}, D, K_\mu$ along with the supersymmetry generator $Q_a$ and the special superconformal generator $S_a$. The Lie super-algebra that these generators obey as well as their action on primary super fields are provided in appendix \ref{sec:algbera}. We provide our notation in appendix \ref{sec:notation}. 
 
 By performing a Fourier transform with respect to the $x^\mu$, we end up in momentum superspace which is described by the coordinates $(p_\mu,\theta^a)$, where $p_\mu$ is the three-momentum and $\theta^a$ are the same Grassmann coordinates as in the position superspace.
 
Let us now study SCFT in this superspace formalism. The quantities that we are interested in obtaining are correlation functions involving primary superfields. In particular the superfields that we are interested in are symmetric traceless conserved super-currents, which we shall henceforth just refer to as a supercurrent for brevity. A spin $s$ supercurrent has the following component expansion in the superspace \cite{Nizami:2013tpa}:
\begin{align}\label{Jsintheta}
    &\mathbf{J}_{s}^{a_1\cdots a_{2s}}(\theta,\mathbf{x})=J_{s}^{a_1\cdots a_{2s}}(\mathbf{x})+\theta_{m}J_{s+\frac{1}{2}}^{(a_1\cdots a_{2s_1}m)}(\mathbf{x})-i\frac{\theta^2}{4}(\slashed{\partial})^{a_1}_{m}J_{s}^{a_2\cdots a_{2s}m}(\mathbf{x}).
\end{align}
The indices $a_1\cdots a_{2s}$ are symmetrized. $J_s$ and $J_{s+\frac{1}{2}}$ are component currents that are conserved. The supercurrent \eqref{Jsintheta} satisfies the following conservation equation:
\begin{align}\label{conservation}
    D_{a_1}\mathbf{J}_{s}^{a_1\cdots a_{2s}}(\theta,\mathbf{x})=0,
\end{align}
where $D_{a_1}=\frac{\partial}{\partial\theta^{a_1}}-\frac{i}{2}\theta_b (\sigma^\mu)^b_{a_1} \partial_\mu$, is the supercovariant derivative.\\
Performing a fourier transform with respect $\mathbf{x}$, we obtain the momentum superspace counterpart to \eqref{Jsintheta}:
\begin{align}\label{Jsinthetamomspace}
    &\mathbf{J}_{s}^{a_1\cdots a_{2s}}(\theta,\mathbf{p})=J_{s}^{a_1\cdots a_{2s}}(\mathbf{p})+\theta_{m}J_{s+\frac{1}{2}}^{(a_1\cdots a_{2s_1}m)}(\mathbf{p})+\frac{\theta^2}{4}(\slashed{p})^{a_1}_{m}J_{s}^{a_2\cdots a_{2s}m}(\mathbf{p}).
\end{align}
We can now construct correlation functions of these operators using the definition \eqref{Jsinthetamomspace}. The super-correlators obey the $\mathcal{N}=1$ super ward identities. For instance, consider the supersymmetry generator $Q_a$. It has the following action on superfields:
\begin{align}
    Q_{ia}=\left(\frac{\partial}{\partial\theta_i^a}-\frac{(\slashed p_i)_a^b}{2}\theta_{ib}\right).
\end{align}
For an n point correlation function, invariance under the Q supersymmetry demands that,
\begin{align}
    \sum_{i=1}^{n} Q_{ia}\langle \mathbf{J}_{s_1}^{a_1\cdots a_{2s_1}}(\theta_1,\mathbf{p}_1) \dots \mathbf{J}_{s_n}^{c_1\cdots c_{2s_n}}(\theta_n,\mathbf{p}_n)\rangle=0.  
\end{align}
This implies that n point correlators take the following form \cite{Inbasekar:2015tsa,Aharony:2019mbc}:
\begin{align}\label{thetaexponential}
  \notag \langle \mathbf{J}_{s_1}^{a_1\cdots a_{2s_1}}(\theta_1,\mathbf{p}_1) \dots \mathbf{J}_{s_n}^{c_1\cdots c_{2s_n}}(\theta_n,\mathbf{p}_n)\rangle&=e^{-\frac{1}{2n}\left(\theta_1^a+\cdots+\theta_n^a\right)\left(\theta_{1b}(\slashed p_1)_{ab}+\cdots+\theta_{nb}(\slashed p_n)_{ab}\right)}\\ &\qquad\qquad F^{a_1\cdots c_{2s_n}}(\{\theta_i-\theta_j\},\{\mathbf{p}_i\}),
\end{align}
where,
\begin{align}\label{functiontosolve}
    F^{a_1\cdots c_{2s_n}}(\{\theta_i-\theta_j\},\{\mathbf{p}_i\})=F_1^{a_1\cdots c_{2s_n}}(\{\mathbf{p}_i\})+(\theta_{1m}-\theta_{2m})F_2^{a_1\cdots c_{2s_n}m}(\{\mathbf{p}_i\})+\cdots,
\end{align}
contains undetermined functions $F_i(\{\mathbf{p}_i\})$ packaged together via a Grassmann spinor expansion.
One can now impose the Ward identities due to the other superconformal generators and constrain the form of these functions. These constraints, however, take the form of coupled partial differential equations involving the $F_i(\{\mathbf{p}_i\})$ and hence quite difficult to solve. There is, however, an alternative method to proceed. We expand the superfields in the correlator using \eqref{Jsinthetamomspace}. The result is a sum of component correlators arranged in a Grassmann spinor expansion. The invariance of the individual component correlators guarantees invariance under the conformal transformations. Please see appendix \ref{sec:KsusygivesK} for proof. We then substitute this component correlator expansion into the LHS of \eqref{thetaexponential} and equate it to the RHS order by order in the Grassmann spinor expansion. Once the resulting \textit{algebraic} equations involving the component correlators and the $F_i(\{\mathbf{p}_i\})$ are solved, the resulting quantity is invariant under the action of the entire superconformal algebra.
\subsection{Constraining super correlators in momentum super space}
Let us now illustrate the methodology that we just outlined via a three-point example. Consider the spins $s_1=\frac{3}{2},s_2=s_3=\frac{1}{2}$. Using the superfield expansions provided in \eqref{Jsinthetamomspace} we obtain,
\begin{align}\label{J3halfJhalfJhalfcomponentexp}
    &\langle\mathbf{J_{\frac{3}{2}}}^{(efg)}\mathbf{J_{\frac{1}{2}}}^a\mathbf{J_{\frac{1}{2}}}^b\rangle=\theta_{1h}\langle T^{(efgh)}O_{\frac{1}{2}}^a O_{\frac{1}{2}}^b\rangle-\theta_{2c}\langle J_{\frac{3}{2}}^{(efg)}J^{(ac)}O_{\frac{1}{2}}^b\rangle+\theta_{3d}\langle J_{\frac{3}{2}}^{(efg)}O_{\frac{1}{2}}^a J^{(bd)}\rangle\notag\\
    &\quad\quad+\frac{1}{4}\theta_{1h}\theta_2^2(\slashed p_2)^a_c \langle T^{(efgh)}O_{\frac{1}{2}}^c O_{\frac{1}{2}}^b\rangle+\frac{1}{4}\theta_{1h}\theta_3^2(\slashed p_3)^b_d\langle T^{(efgh)}O_{\frac{1}{2}}^a O_{\frac{1}{2}}^d\rangle-\frac{1}{4}\theta_{2c}\theta_1^2\langle J_{\frac{3}{2}}^{(hfg)}J^{(ac)}O_{\frac{1}{2}}^b\rangle\notag\\
    &\qquad-\frac{1}{4}\theta_{2c}\theta_3^2(\slashed p_3)^b_d\langle J_{\frac{3}{2}}^{(efg)}J^{(ac)}O_{\frac{1}{2}}^d\rangle+\frac{1}{4}\theta_{3d}\theta_1^2(\slashed p_1)^e_h \langle J_{\frac{3}{2}}^{(hfg)}O_{\frac{1}{2}}^a J^{(bd)}\rangle+\frac{1}{4}\theta_{3d}\theta_2^2(\slashed p_2)^a_c\langle J_{\frac{3}{2}}^{(efg)}O_{\frac{1}{2}}^c J^{bd)}\rangle\notag\\
&\qquad+ \theta_{1h}\theta_{2c}\theta_{3d}\langle T^{(efgh)}J^{(ac)}J^{(bd)}\rangle+\frac{1}{16}\theta_1^2\theta_2^2\theta_{3d}(\slashed p_1)^e_h (\slashed p_2)^a_c\langle J_{\frac{3}{2}}^{(hfg)}O_{\frac{1}{2}}^c J^{(bd)}\rangle\notag\\
&\qquad+\frac{1}{16}\theta_1^2\theta_3^2\theta_{2c}(\slashed p_1)^e_h(\slashed p_3)^b_d\langle J_{\frac{3}{2}}^{(hfg)}J^{(ac)}O_{\frac{1}{2}}^d\rangle+\frac{1}{16}\theta_2^2\theta_3^2 \theta_{1h}(\slashed p_2)^a_c(\slashed p_3)^b_d\langle T^{(efgh)}O_{\frac{1}{2}}^c O_{\frac{1}{2}}^d\rangle,
\end{align}
 while the analogue of \eqref{thetaexponential} is given by,
 \begin{align}\label{J3halfJhalfJhalfexponential}
    &\langle\mathbf{J_{\frac{3}{2}}}^{(efg)}\mathbf{J_{\frac{1}{2}}}^a\mathbf{J_{\frac{1}{2}}}^b\rangle=e^{\frac{1}{6}(\theta_1^m+\theta_2^m+\theta_3^m)\left((\theta_1-\theta_2)^n (\slashed p_2)_{mn}+(\theta_1-\theta_3)^n (\slashed p_3)_{mn}\right)}\Big( A_1^{(efg)abl} (\theta_{1}-\theta_{2})_l\notag\\
    &\qquad\quad+A_2^{(efg)abl}(\theta_{1}-\theta_{3})_l+B_{1}^{(efg)abl}(\theta_{1}-\theta_{2})_l(\theta_1-\theta_3)^2+B_2^{(efg)abl}(\theta_{1}-\theta_{3})_l(\theta_1-\theta_2)^2\Big).
\end{align}
We now equate equations \eqref{J3halfJhalfJhalfcomponentexp} and \eqref{J3halfJhalfJhalfexponential} order by order in the Grassmann spinor expansion.  For instance, we obtain at the lowest order, 
\begin{align}
    &A_1^{(efg)abl}=\langle J_{3/2}^{(efg)}J^{(al)}O_{1/2}^b\rangle,\notag\\
    &A_2^{(efg)abl}=-\langle J_{3/2}^{(efg)}O_{1/2}^a J^{(bl)}\rangle,\notag\\
    &A_1^{(efg)abl}+A_2^{(efg)abl}=\langle T^{(efgl)}O_{1/2}^a O_{1/2}^b\rangle,
\end{align}
which forces the relation,
\begin{align}\label{TOhOhvsJ3by2OhJ}
    \langle T^{(efgl)}O_{1/2}^a O_{1/2}^b\rangle=\langle J_{3/2}^{(efg)}J^{(al)}O_{1/2}^b\rangle-\langle J_{3/2}^{(efg)}O_{1/2}^a J^{(bl)}\rangle.
\end{align}
One can independently check that this holds true by computing LHS and RHS seperately. Going to higher orders in the Grasmann expansion yields similar but albeit more complicated and not very insightful relations also involving the component $\langle TJJ\rangle$ correlator.

In order to proceed further, we then use the fact that in a generic 3d CFT, a three point (component) correlator can have at most three independent structures \cite{Giombi:2011rz}:
\begin{align}\label{s1s2s3comp}
    \langle J_{s_1}J_{s_2}J_{s_3}\rangle=n_b \langle J_{s_1}J_{s_2}J_{s_3}\rangle_{b}+n_f \langle J_{s_1}J_{s_2}J_{s_3}\rangle_{f}+n_{odd}\langle J_{s_1}J_{s_2}J_{s_3}\rangle_{odd},
\end{align}
where the first and the second term are the free bosonic and free fermionic correlators whilst the third term is a parity odd piece.
Using \eqref{s1s2s3comp} for each of the component correlators appearing in the superfield expansion and demanding that relations such as \eqref{TOhOhvsJ3by2OhJ} are satisfied, reduces the number of independent coeffcients in \eqref{s1s2s3comp} for every component correlator. In fact, for the correlator in \eqref{J3halfJhalfJhalfcomponentexp}, we find $n_f=n_b$ and $n_{odd}=0$ for all the component correlators thereby obtaining a single parity even solution:
\begin{align}\label{J3by2JhJhComps}
    \langle TO_{1/2}O_{1/2}\rangle&=n_b\big(\langle TO_{1/2}O_{1/2}\rangle_{FB}+\langle TO_{1/2}O_{1/2}\rangle_{FF}\big),\notag\\
    \langle J_{3/2}J O_{1/2}\rangle&=n_b\big(\langle J_{3/2}J O_{1/2}\rangle_{FB}+\langle J_{3/2}J O_{1/2}\rangle_{FF}\big),\notag\\
    \langle TJJ\rangle&=n_b\big(\langle TJJ\rangle_{FB}+\langle TJJ\rangle_{FF}\rangle\big),
\end{align}
a result that is consistent with \cite{Buchbinder:2023fqv}, that is, the existence of a single parity even solution. The final expression for the correlator is obtained by substituting \eqref{J3by2JhJhComps}  back into \eqref{J3halfJhalfJhalfcomponentexp} or equivalently \eqref{J3halfJhalfJhalfexponential}. By construction this correlator is invariant under the action of all the generators of the superconformal algebra.

While we were systematically able to solve the equations at every order in the Grassmann spinor expansion, the procedure is quite complicated. Other than dealing with various degeneracy identities\footnote{An example of such an identity is $(\slashed p_1)^l_f (\slashed p_2)^{mb}=(\slashed p_1)^{bl} (\slashed p_2)^m_f-(\slashed p_1)^{kl} (\slashed p_2)^m_k \delta^b_f$.}, there is no simple generalization to arbitrary spin correlators, no obvious connection to the flat space scattering amplitudes in four dimensions, or the existence of double copy relations between various super-correlators. Hence, we define our new variables in the next section to surpass this tedious process. We give a worked-out example of the same correlator and get the same constraints on the coefficients using our new formalism in appendix \ref{sec:example}, circumventing all the complexities of momentum super-space variables in one go.
\subsection{En route to Spinor Helicity and Grassmann Twistor Variables}
Rather than attempt a tour de force and try to obtain correlators the above way, we discovered a different set of variables that simplify the analysis significantly. As a first step, let us use spinor helicity variable defined as 
\begin{align}\label{pmuspinorhelicity}
    &p_\mu=\frac{1}{2}(\sigma^\mu)^a_b\lambda_a\,\overline{\lambda}^b,
\end{align}
where $\lambda$ and $\,\overline{\lambda}$ are two component commuting spinors. The momentum $p_\mu$ is invariant under the little group transformation, $\lambda\to r \lambda,\,\overline{\lambda}\to r^{-1}\,\overline{\lambda}$, $r\in \mathbb{C}$. As in the case of non-supersymmetruc correlators, the introduction of spinor helicity variables greatly helps us in dealing with degeneracies which leads to simplification of algebra as well as expressions. However, to truly exploit the power of spinor helicity variables, we also need to express the Grassmann spinors $\theta^a$ in the basis of $\lambda$ and $\,\overline{\lambda}$. We define,
\begin{equation}\label{etavariables}
\theta^a=\frac{\,\overline{\eta}\lambda^a+\eta \,\overline{\lambda}^a}{2p},
\end{equation}
where $\eta$ is a complex Grassmann variable and $\,\overline{\eta}$ is it's complex conjugate.
 This definition is consistent with the fact that the dimensionality of $\theta$ is $-\frac{1}{2}$. Further, $\eta$ and $\,\overline{\eta}$ must transform as $\eta\to r\eta,\,\overline{\eta}\to r^{-1}\,\overline{\eta}$ under little group scalings so that $\theta$ remains unaffected.\\
 We then contract the superfield \eqref{Jsinthetamomspace} with the polarization spinors which are given by,
 \begin{align}
    &\zeta_{a}^{-}=\frac{\lambda_a}{\sqrt{p}},~~~~\zeta_{a}^{+}=\frac{\,\overline{\lambda}_a}{\sqrt{p}},
\end{align}
and express the grassmann spinor $\theta^a$ in terms of the $\eta,\,\overline{\eta}$ variables defined in \eqref{etavariables}. We obtain the super-current in the $\pm s$ helicities\footnote{We also use the fact that the supercurrent \eqref{Jsinthetamomspace} is symmetric, traceless and conserved and hence it has only two independent components which are the $h=\pm s$ ones.} ,
\begin{align}\label{Jsineta}
    &\mathbf{J}_s^{-}=e^{-\frac{\eta \,\overline{\eta}}{4}}J_s^{-}+\frac{\,\overline{\eta}}{2\sqrt{p}}J_{s+\frac{1}{2}}^{-},\notag\\
    &\mathbf{J}_s^{+}=e^{\frac{\eta \,\overline{\eta}}{4}}J_s^{+}+\frac{\eta}{2\sqrt{p}}J_{s+\frac{1}{2}}^{+},
\end{align}
where, $\mathbf{J}_s^{\pm}:=\zeta_{a_1}^{\pm}\cdots \zeta_{a_{2s}}^{\pm}\mathbf{J}_{s}^{a_1\cdots a_{2s}}$.
The Super-correlator \eqref{J3halfJhalfJhalfcomponentexp} in these variables becomes extremely simple. For instance in the $(- - +)$ helicity configuration we obtain\footnote{Similar expressions can be obtained for all other helicities.},
\begin{align}
    \langle \mathbf{J}_{\frac{3}{2}}^{-}\mathbf{J}_{\frac{1}{2}}^{-}\mathbf{J}_{\frac{1}{2}}^{+}\rangle&=\frac{1}{2 \sqrt{p_3}}\langle J^-_{3/2}O^-_{1/2}T^+\rangle \bigg( e^{-\frac{\eta_1\bar{\eta}_1}{4}} e^{-\frac{\eta_2\bar{\eta}_2}{4}} \eta_3\notag\\
    &\quad\qquad\quad -\frac{1}{E}\left(e^{-\frac{\eta_1\bar{\eta}_1}{4}} e^{\frac{\eta_3\bar{\eta}_3}{4}} \bar{\eta}_2 \langle 23\rangle-e^{-\frac{\eta_2\bar{\eta}_2}{4}} e^{\frac{\eta_3\bar{\eta}_3}{4}} \bar{\eta}_1 \langle 31\rangle-\frac{\Bar{\eta}_1\Bar{\eta}_2\eta_3}{2}\langle 12\rangle\right)\bigg).
 \end{align}
Further, the exponential of the Grassmann bilinears that appear in this expression suggests us to perform a Grassmann ``half" Fourier transform analogous to the twistor transform \cite{Witten:2003nn,Elvang:2013cua}.
Given a function $F(\eta,\,\overline{\eta})$ we define the Grassmann Twistor transform as follows:
\begin{align}\label{GrassmannTwistorTransform}
    \Tilde{F}(\eta,\chi):=\int d\,\overline{\eta}~e^{-\frac{\chi\,\overline{\eta}}{4}}F(\eta,\,\overline{\eta}).
\end{align}
We then make a variable change from $(\eta,\chi)$ to $(\xi_+,\xi_-)$ which are defined as,
\begin{align}\label{xipmdef}
    \xi_{\pm}=\chi\pm \eta.
\end{align}
In these new Grassmann ``Twistor" Variables, the correlator takes the following beautiful form\footnote{We can then proceed to obtain similar expressions in all the other helicity configurations as well.}:
\begin{align}\label{J3byJhJhmmpXi}
    \langle\tilde{\mathbf{J}}^-_{\frac{3}{2}} \tilde{\mathbf{J}}^-_{\frac{1}{2}} \tilde{\mathbf{J}}^+_{\frac{1}{2}}\rangle&=-\frac{\xi_{3-} }{256\sqrt{p_3}}\,\langle J^-_{\frac{3}{2}}J^-_{\frac{1}{2}}J^+_{1}\rangle\,\left[\xi_{1+}\xi_{2+}\xi_{3+} -\frac{8}{E}\big(\xi_{1+}\langle 23 \rangle+\xi_{2+}\langle 31 \rangle+\xi_{3+}\langle 12 \rangle
\big)\right].
\end{align}
 Contrasted with \eqref{J3halfJhalfJhalfcomponentexp}, the new representation \eqref{J3byJhJhmmpXi} is not only simpler but also appears in a language that is homologous to the four-dimensional flat space scattering amplitudes \cite{Elvang:2013cua}. Further, we shall see in the next section that the structure appearing inside the brackets in \eqref{J3byJhJhmmpXi} is one of two universal structures that appear in three-point super-correlators with arbitrary (half) integer spin insertions! Motivated by all this simplicity, let us, from first principles, develop our formalism in these new variables.
 \subsection*{Grassmann Twistor Variables}
The superfield expansion in the grassmann twistor variable can be obtained by performing the transformation \eqref{GrassmannTwistorTransform} on \eqref{Jsineta}. The result is \footnote{We note the similarity between \eqref{JsinXipm} and the $\mathcal{N}=1$ superfields in the four dimensional flat space literature, see for instance \cite{Elvang:2011fx}.},
\begin{align}\label{JsinXipm}
     &\mathbf{\tilde{J}}_s^{-}=\frac{1}{4}\bigg(\xi_{+}J_{s}^{-}+\frac{2}{\sqrt{p}}J_{s+\frac{1}{2}}^{-}\bigg),\notag\\
    &\mathbf{\tilde{J}}_s^{+}=\frac{\xi_-}{4}\bigg(J_{s}^{+}+\frac{\xi_+}{4\sqrt{p}}J_{s+\frac{1}{2}}^+\bigg).
\end{align}
We then construct correlation functions of the super-currents in these new variables using the superfield expansion \eqref{JsinXipm}. The only Ward identity that we need to impose is the one due to the $Q$ supersymmetry. In these new variables we have,
\begin{align}\label{Qop}
    Q_{ia}=2 \lambda_{ia}\frac{\partial}{\partial \xi_{i+}}+\frac{\,\overline{\lambda}_{ia}}{4}\xi_{i+}.
\end{align}
The associated Ward identity reads,
\begin{align}\label{QWardIdentity}
    \sum_{i=1}^{n}Q_{ia}\langle \mathbf{\tilde{J}}_{s_1}^{\pm}\cdots \mathbf{\tilde{J}}_{s_n}^{\pm}\rangle=0.
\end{align}
As we can see from \eqref{Qop}, $Q_a=\displaystyle\sum_{i=1}^{n}Q_{ia}$ is a two component spinor operator. Therefore, it can be written in the following way\footnote{We choose to work in the $(\lambda_{1a}$, $\Bar{\lambda}_{1a})$ basis. One can also choose to work in other basis but the results that we obtain will be basis independent.}:
\begin{equation}
Q_a=\lambda_{1a}q+\,\overline{\lambda}_{1a}\,\overline{q},
\end{equation}
for some $q,\,\overline{q}$. It is easy to show (using the Schouten identity, see appendix \ref{sec:notation}) that we have,
\begin{align}
    q&=2\frac{\partial}{\partial\xi_{1+}}+\frac{1}{p_1}\sum_{i=2}^{n}\bigg(\langle\overline{1}i\rangle\frac{\partial}{\partial\xi_{i+}}+\frac{\langle\overline{1}\,\overline{i}\rangle}{8}\xi_{i+}\bigg),\notag\\
    \,\overline{q}&=\frac{\xi_{1+}}{4}-\frac{1}{p_1}\sum_{i=2}^{n}\bigg(\langle 1 i\rangle\frac{\partial}{\partial \xi_{i+}}+\frac{\langle 1\,\overline{i}\rangle}{8}\xi_{i+}\bigg).
\end{align}
Thus, the $Q$ Ward identity \eqref{QWardIdentity}, splits into two simpler Ward identities viz,
\begin{align}\label{qqbarWardId}
    q\langle \mathbf{\tilde{J}}_{s_1}^{\pm}\cdots \mathbf{\tilde{J}}_{s_n}^{\pm}\rangle&=0,\notag\\
    \overline{q}\langle \mathbf{\tilde{J}}_{s_1}^{\pm}\cdots \mathbf{\tilde{J}}_{s_n}^{\pm}\rangle&=0.
\end{align}
We need not impose the Ward identities at the level of the Super-correlator for the same reason that we mentioned just below \eqref{functiontosolve}, i.e., imposing conformal invariance at the component level suffices to have conformal invariance at the level of the Super-correlator. Further, the different coefficients that appear in the component correlators \eqref{s1s2s3comp} get related to each other due to the $Q_a$ ward identities \eqref{qqbarWardId}.
\section{Correlation functions in $\mathcal{N}=1$ SCFTs}\label{sec:Neq1Corr}
In this section, we shall present our final results for two and three point correlators in the $\xi_{\pm}$ variables \eqref{xipmdef}. 
\subsection{Two Point Functions}
\subsubsection{$\langle \mathbf{\tilde{J}}_{s}\mathbf{\tilde{J}}_s\rangle$,~$s\in\mathbb{Z}_{>0}$}
We obtain the following expressions for the correlator in the $(- -)$ and $(+ +)$ helicity configurations:
\begin{align}\label{twopointintspin}
    &\langle \mathbf{\tilde{J}}^-_{s}\mathbf{\tilde{J}}^-_s\rangle=\frac{\langle 1 2\rangle^{2s}}{16 p_1}\bigg(\xi_{1+}\xi_{2+}-\frac{4\langle 1 2\rangle}{p_1}\bigg),\notag\\
    &\langle \mathbf{\tilde{J}}^+_{s}\mathbf{\tilde{J}}^+_s\rangle=-\xi_{1-}\xi_{2-}\frac{\langle\overline{1}\,\overline{2}\rangle^{2s+1}}{256 p_1^2}\bigg(\xi_{1+}\xi_{2+}-\frac{4\langle 1 2\rangle}{p_1}\bigg).
\end{align}
\subsubsection{$\langle \mathbf{\tilde{J}}_{s}\mathbf{\tilde{J}}_s\rangle$,~$s=k+\frac{1}{2},~k\in\mathbb{Z}_{\ge 0}$}
The expressions of the correlator in the $(- -)$ and $(+ +)$ helicity configurations are given by,
\begin{align}\label{twopointhalfintspin}
    &\langle \mathbf{\tilde{J}}^-_{s}\mathbf{\tilde{J}}^-_s\rangle=-\frac{\langle 1 2\rangle^{2s}}{16 p_1}\bigg(\xi_{1+}\xi_{2+}-\frac{4\langle 1 2\rangle}{p_1}\bigg),\notag\\
    &\langle \mathbf{\tilde{J}}^+_{s}\mathbf{\tilde{J}}^+_s\rangle=\xi_{1-}\xi_{2-}\frac{\langle\overline{1}\,\overline{2}\rangle^{2s+1}}{256 p_1^2}\bigg(\xi_{1+}\xi_{2+}-\frac{4\langle 1 2\rangle}{p_1}\bigg).
\end{align}
\subsubsection{Summary}
Based on our results, \eqref{twopointintspin} and \eqref{twopointhalfintspin}, we see that general two point functions of any (half) integer spin $s$ conserved super-currents take the following form:
\begin{align}
    &\langle \mathbf{\tilde{J}}^-_{s}\mathbf{\tilde{J}}^-_s\rangle=(-1)^{2s}\frac{\langle 1 2\rangle^{2s}}{16 p_1}\Xi_2,~~\langle \mathbf{\tilde{J}}^+_{s}\mathbf{\tilde{J}}^+_s\rangle=\xi_{1-}\xi_{2-}(-1)^{2s+1}\frac{\langle\overline{1}\,\overline{2}\rangle^{2s+1}}{256 p_1^2}\Xi_2,
\end{align}
where we have defined the two point building block,
\begin{align}\label{twopointbuildingblock}
    \Xi_2=\bigg(\xi_{1+}\xi_{2+}-\frac{4\langle 1 2\rangle}{p_1}\bigg).
\end{align}
Thus we see that for any spin $s$, the two point correlators are given by a simple kinematic factor times a universal factor $\Xi_2$.
We will find a similar structure at the level of three points as well.

\subsection{Three Point Correlation Functions}
In this subsection, we present our results for three point correlators with arbitrary (half) integer spin insertions. In contrast to the previous analysis in position space \cite{Nizami:2013tpa,Buchbinder:2023fqv} etc..., we find a universal form for any super-correlator independent of the spins of the insertions.
\subsubsection{$\langle \tilde{\mathbf{J}}_{s_1}\tilde{\mathbf{J}}_{s_2}\tilde{\mathbf{J}}_{s_3}\rangle$, $s_1,s_2,s_3\,\in \mathbb{Z}_{> 0}$}
The correlator of three integer spin conserved super-currents is given by the following expressions in the various helicity configurations:
\begin{equation}\label{threepointInt}
  \begin{split}
    \langle \tilde{\mathbf{J}}_{s_1}^-\tilde{\mathbf{J}}_{s_2}^-\tilde{\mathbf{J}}_{s_3}^-\rangle&=\frac{1}{64} \langle J_{s_1}^-J_{s_2}^-J_{s_3}^-\rangle \Gamma_{3},\\
     \langle \tilde{\mathbf{J}}_{s_1}^+\tilde{\mathbf{J}}_{s_2}^+\tilde{\mathbf{J}}_{s_3}^-\rangle&=-\frac{\xi_{1-}\xi_{2-}}{512}\,\frac{\langle J_{s_1}^+J_{s_2}^+J_{s_3}^-\rangle E}{\langle12\rangle} \,\Gamma_{3},\\ 
     \langle \tilde{\mathbf{J}}_{s_1}^+\tilde{\mathbf{J}}_{s_2}^-\tilde{\mathbf{J}}_{s_3}^+\rangle&=-\frac{\xi_{1-}\xi_{3-}}{512}\, \frac{\langle J_{s_1}^+J_{s_2}^-J_{s_3}^+\rangle E}{\langle13\rangle}\,\Gamma_{3},\\
    \langle \tilde{\mathbf{J}}_{s_1}^-\tilde{\mathbf{J}}_{s_2}^+\tilde{\mathbf{J}}_{s_3}^+\rangle&=-\frac{\xi_{2-}\xi_{3-}}{512}\, \frac{\langle J_{s_1}^-J_{s_2}^+J_{s_3}^+\rangle E}{\langle23\rangle}\, \Gamma_{3},
  \end{split}\quad
  \begin{split}
  \langle \tilde{\mathbf{J}}_{s_1}^+\tilde{\mathbf{J}}_{s_2}^+\tilde{\mathbf{J}}_{s_3}^+\rangle&=\frac{\xi_{1-}\xi_{2-}\xi_{3-}}{512}\, \langle J_{s_1}^+J_{s_2}^+J_{s_3}^+\rangle\, \Xi_3,\\
  \langle \tilde{\mathbf{J}}_{s_1}^-\tilde{\mathbf{J}}_{s_2}^-\tilde{\mathbf{J}}_{s_3}^+\rangle&=-\frac{\xi_{3-}}{64}\,  \frac{\langle J_{s_1}^-J_{s_2}^-J_{s_3}^+\rangle E}{\langle\overline{1}\,\overline{2}\rangle}\, \Xi_3,\\
  \langle \tilde{\mathbf{J}}_{s_1}^-\tilde{\mathbf{J}}_{s_2}^+\tilde{\mathbf{J}}_{s_3}^-\rangle&=-\frac{\xi_{2-}}{64}\,  \frac{\langle J_{s_1}^-J_{s_2}^+J_{s_3}^-\rangle E}{\langle\overline{3}\,\overline{1}\rangle} \,\Xi_3,\\
  \langle \tilde{\mathbf{J}}_{s_1}^+\tilde{\mathbf{J}}_{s_2}^-\tilde{\mathbf{J}}_{s_3}^-\rangle&=-\frac{\xi_{1-}}{64}\,\frac{\langle J_{s_1}^+J_{s_2}^-J_{s_3}^-\rangle E}{\langle\overline{2}\,\overline{3}\rangle}\, \Xi_3,
  \end{split}
  \end{equation}  
where $\Gamma_3$ and $\Xi_3$ are given by,
\begin{align}\label{threepointbuildingblocks}
\Gamma_{3}&=\left[\xi_{1+}\xi_{2+}\xi_{3+} -\frac{8}{E}\big(\xi_{1+}\langle 23 \rangle+\xi_{2+}\langle 31 \rangle+\xi_{3+}\langle 12 \rangle
\big)\right],\notag\\
\Xi_3&=\left[8-\frac{1}{E}\big(\xi_{1+}\xi_{2+}\langle\overline{1}\,\overline{2}\rangle+\xi_{2+}\xi_{3+}\langle\overline{2}\,\overline{3}\rangle+\xi_{3+}\xi_{1+}\langle\overline{3}\,\overline{1}\rangle\big)  \right],
\end{align}
 which we note is similar to the two point case \eqref{twopointbuildingblock} where super-correlators are constructed by multiplying a component correlator and a universal spin independent factor, except that in the three point case, we have two such universal factors\footnote{We note that $\Gamma_3$ is antisymmetric under any exchange of $(1,2,3)$ whilst $\Xi_3$ is symmetric under any such exchange.}.
Further, we find that for super-correlators that obey the triangle inequality, the only non zero components are the $(- - -)$ and $(+ + +)$ helicity configurations, i.e, the Super-correlator is homogeneous and has both a parity even and a parity odd solution. In fact we have the universal formulae \cite{Jain:2021whr}.
 \begin{align}
     \langle J_{s_1}^{-}J_{s_2}^{-}J_{s_3}^{-}\rangle&=(c_{even}+i c_{odd})\frac{\langle 1 2\rangle^{s_1+s_2-s_3}\langle 2 3\rangle^{s_2+s_3-s_1}\langle 3 1\rangle^{s_1+s_3-s_2}}{E^{s_1+s_2+s_3}}p_1^{s_1-1}p_2^{s_2-1}p_3^{s_3-1},\notag\\
     \langle J_{s_1}^{+}J_{s_2}^{+}J_{s_3}^{+}\rangle&=(c_{even}-i c_{odd})\frac{\langle \Bar{1} \Bar{2}\rangle^{s_1+s_2-s_3}\langle \Bar{2} \Bar{3}\rangle^{s_2+s_3-s_1}\langle \Bar{3} \Bar{1}\rangle^{s_1+s_3-s_2}}{E^{s_1+s_2+s_3}}p_1^{s_1-1}p_2^{s_2-1}p_3^{s_3-1},
 \end{align}
 where $c_{even}$ and $c_{odd}$ are the OPE coefficients corresponding to the parity even and parity odd structures respectively.
The relations between the various component correlators as in \eqref{threepointInt} relate the various component correlators to each other, thereby fixing their OPE coeffcients just in terms of $c_{even}$ and $c_{odd}$.

For super-correlators that violate the triangle inequality, their is no parity odd solution as the component correlator on the RHS of \eqref{threepointInt} has no parity odd piece in this case. There exists however, an even piece. In terms of the free bosonic and free fermionic correlators it is given by,
\begin{align}\label{s1s2s3intoutsidetriangle}
    \langle J_{s_1}^{\pm}J_{s_2}^{\pm}J_{s_3}^{\pm}\rangle=n_b\bigg(\langle J_{s_1}^{\pm}J_{s_2}^{\pm}J_{s_3}^{\pm}\rangle_{FB}-\langle J_{s_1}^{\pm}J_{s_2}^{\pm}J_{s_3}^{\pm}\rangle_{FF}\bigg).
\end{align}
Substituting \eqref{s1s2s3intoutsidetriangle} in \eqref{threepointInt} and reading off the component correlators using \eqref{JsinXipm}, the other component correlators can be obtained.
 \subsubsection{$\langle \tilde{\mathbf{J}}_{s_1}\tilde{\mathbf{J}}_{s_2}\tilde{\mathbf{J}}_{s_3}\rangle$, $s_1,s_2\,\in \mathbb{Z}_{> 0},s_3=k_3+\frac{1}{2},~k_3\in \mathbb{Z}_{\ge 0}$}
 We obtained the following expressions for the various helicity configurations for this half integer spin correlator:
\begin{equation}
\label{threepoint2Int1halfInt}
    \begin{split}
\langle\tilde{\mathbf{J}}^-_{s_1} \tilde{\mathbf{J}}^-_{s_2} \tilde{\mathbf{J}}^-_{s_3}\rangle&=-\frac{ 1}{32}\frac{\langle J^-_{s_1}J^-_{s_2}J^-_{k_3+1}\rangle E}{\langle\overline{1}\,\overline{2}\rangle\sqrt{p_3}}\,\,\Xi_3,\\
\langle\tilde{\mathbf{J}}^+_{s_1} \tilde{\mathbf{J}}^+_{s_2} \tilde{\mathbf{J}}^-_{s_3}\rangle&=\frac{\xi_{1-}\xi_{2-} }{256}\frac{\langle J^+_{s_1}J^+_{s_2}J^-_{k_3+1}\rangle}{\sqrt{p_3}}\,\Xi_3,\\  
\langle\tilde{\mathbf{J}}^+_{s_1} \tilde{\mathbf{J}}^-_{s_2} \tilde{\mathbf{J}}^+_{s_3}\rangle&=\frac{\xi_{1-}\xi_{3-} }{256}\,\frac{\langle J^+_{s_1}J^-_{s_2}J^+_{k_3+1}\rangle E}{\langle\overline{2}\,\overline{3}\rangle\sqrt{p_3}}\,\Xi_3,\\
\langle\tilde{\mathbf{J}}^-_{s_1} \tilde{\mathbf{J}}^+_{s_2} \tilde{\mathbf{J}}^+_{s_3}\rangle&=\frac{\xi_{2-}\xi_{3-} }{256}\frac{\langle J^-_{s_1}J^+_{s_2}J^+_{k_3+1}\rangle E}{\langle\overline{3}\,\overline{1}\rangle\sqrt{p_3}}\,\,\Xi_3,
    \end{split}\quad
    \begin{split}
\langle\tilde{\mathbf{J}}^+_{s_1} \tilde{\mathbf{J}}^+_{s_2} \tilde{\mathbf{J}}^+_{s_3}\rangle&=-\frac{\xi_{1-}\xi_{2-}\xi_{3-} }{2048}\frac{\langle J^+_{s_1}J^+_{s_2}J^+_{k_3+1}\rangle E}{\langle12\rangle\sqrt{p_3}}\,\,\Gamma_3,\\
\langle\tilde{\mathbf{J}}^-_{s_1} \tilde{\mathbf{J}}^-_{s_2} \tilde{\mathbf{J}}^+_{s_3}\rangle&=\frac{\xi_{3-} }{256}\,\frac{\langle J^-_{s_1}J^-_{s_2}J^+_{k_3+1}\rangle}{\sqrt{p_3}}\,\Gamma_3,\\
\langle\tilde{\mathbf{J}}^-_{s_1} \tilde{\mathbf{J}}^+_{s_2} \tilde{\mathbf{J}}^-_{s_3}\rangle&=\frac{\xi_{2-} }{256}\,\frac{\langle J^-_{s_1}J^+_{s_2}J^-_{k_3+1}\rangle E}{\langle23\rangle\sqrt{p_3}}\,\Gamma_3,\\
\langle\tilde{\mathbf{J}}^+_{s_1} \tilde{\mathbf{J}}^-_{s_2} \tilde{\mathbf{J}}^-_{s_3}\rangle&=-\frac{\xi_{1-} }{256}\,\frac{\langle J^+_{s_1}J^-_{s_2}J^-_{k_3+1}\rangle E}{\langle31\rangle\sqrt{p_3}}\,\Gamma_3.
    \end{split}
\end{equation}
In contrast to the previous case with all integer spins, this class of correlators is non-homogeneous, that is, the component correlators appear as a sum of the bosonic and fermionic correlators $\langle~\rangle_{B}+\langle ~\rangle_{F}$. The parity odd structure does not exist for these correlators.
 \subsubsection{$\langle \tilde{\mathbf{J}}_{s_1}\tilde{\mathbf{J}}_{s_2}\tilde{\mathbf{J}}_{s_3}\rangle$, $s_1\in \mathbb{Z}_{> 0}, s_i=k_i+\frac{1}{2}, i=2,3~\text{and}~k_i\in\mathbb{Z}_{\ge 0}$}
 This integer spin super-correlator has the following components:
\begin{equation}\label{threepoint1Int2halfInt}
  \begin{split}
    \langle \tilde{\mathbf{J}}_{s_1}^-\tilde{\mathbf{J}}_{s_2}^-\tilde{\mathbf{J}}_{s_3}^-\rangle&=\frac{1}{64} \langle J_{s_1}^-J_{k_2+\frac{1}{2}}^-J_{k_3+\frac{1}{2}}^-\rangle \Gamma_{3},\\
     \langle \tilde{\mathbf{J}}_{s_1}^+\tilde{\mathbf{J}}_{s_2}^+\tilde{\mathbf{J}}_{s_3}^-\rangle&=\frac{\xi_{1-}\xi_{2-}}{512}\,\frac{\langle J_{s_1}^+J_{k_2+\frac{1}{2}}^+J_{k_3+\frac{1}{2}}^-\rangle E}{\langle12\rangle} \,\Gamma_{3},\\ 
     \langle \tilde{\mathbf{J}}_{s_1}^+\tilde{\mathbf{J}}_{s_2}^-\tilde{\mathbf{J}}_{s_3}^+\rangle&=\frac{\xi_{1-}\xi_{3-}}{512}\, \frac{\langle J_{s_1}^+J_{k_2+\frac{1}{2}}^-J_{k_3+\frac{1}{2}}^+\rangle E}{\langle13\rangle}\,\Gamma_{3},\\
    \langle \tilde{\mathbf{J}}_{s_1}^-\tilde{\mathbf{J}}_{s_2}^+\tilde{\mathbf{J}}_{s_3}^+\rangle&=\frac{\xi_{2-}\xi_{3-}}{512}\, \frac{\langle J_{s_1}^-J_{k_2+\frac{1}{2}}^+J_{k_3+\frac{1}{2}}^+\rangle E}{\langle23\rangle}\, \Gamma_{3},
  \end{split}\quad
  \begin{split}
  \langle \tilde{\mathbf{J}}_{s_1}^+\tilde{\mathbf{J}}_{s_2}^+\tilde{\mathbf{J}}_{s_3}^+\rangle&=-\frac{\xi_{1-}\xi_{2-}\xi_{3-}}{512}\, \langle J_{s_1}^+J_{k_2+\frac{1}{2}}^+J_{k_3+\frac{1}{2}}^+\rangle\, \Xi_3,\\
  \langle \tilde{\mathbf{J}}_{s_1}^-\tilde{\mathbf{J}}_{s_2}^-\tilde{\mathbf{J}}_{s_3}^+\rangle&=\frac{\xi_{3-}}{64}\,  \frac{\langle J_{s_1}^-J_{k_2+\frac{1}{2}}^-J_{k_3+\frac{1}{2}}^+\rangle E}{\langle\overline{1}\,\overline{2}\rangle}\, \Xi_3,\\
  \langle \tilde{\mathbf{J}}_{s_1}^-\tilde{\mathbf{J}}_{s_2}^+\tilde{\mathbf{J}}_{s_3}^-\rangle&=\frac{\xi_{2-}}{64}\,  \frac{\langle J_{s_1}^-J_{k_2+\frac{1}{2}}^+J_{k_3+\frac{1}{2}}^-\rangle E}{\langle\overline{3}\,\overline{1}\rangle} \,\Xi_3,\\
  \langle \tilde{\mathbf{J}}_{s_1}^+\tilde{\mathbf{J}}_{s_2}^-\tilde{\mathbf{J}}_{s_3}^-\rangle&=\frac{\xi_{1-}}{64}\,\frac{\langle J_{s_1}^+J_{k_2+\frac{1}{2}}^-J_{k_3+\frac{1}{2}}^-\rangle E}{\langle\overline{2}\,\overline{3}\rangle}\, \Xi_3.
  \end{split}
  \end{equation}
  Similar to the all integer spin case \eqref{threepointInt}, this correlator also has components that appear as a difference between the bosonic and fermionic structures. In addition, there also exists a parity odd structure when the triangle inequality is satisfied.

 \subsubsection{$\langle \tilde{\mathbf{J}}_{s_1}\tilde{\mathbf{J}}_{s_2}\tilde{\mathbf{J}}_{s_3}\rangle$, $s_i=k_i+\frac{1}{2}, i=1,2,3~\text{and}~k_i\in\mathbb{Z}_{\ge 0}$}
 This correlator involving three half integer insertions has the following expressions in the various helicities:
\begin{equation}\label{threepoint3halfInt}
      \begin{split}
\langle\tilde{\mathbf{J}}^-_{s_1} \tilde{\mathbf{J}}^-_{s_2} \tilde{\mathbf{J}}^-_{s_3}\rangle&=\frac{1}{32}\frac{\langle J^-_{k_1+\frac{1}{2}}J^-_{k_2+\frac{1}{2}}J^-_{k_3+1}\rangle E}{\langle\overline{1}\,\overline{2}\rangle\sqrt{p_3}}\,\,\Xi_3,\\
\langle\tilde{\mathbf{J}}^+_{s_1} \tilde{\mathbf{J}}^+_{s_2} \tilde{\mathbf{J}}^-_{s_3}\rangle&=-\frac{\xi_{1-}\xi_{2-} }{256}\frac{\langle J^+_{k_1+\frac{1}{2}}J^+_{k_2+\frac{1}{2}}J^-_{k_3+1}\rangle}{\sqrt{p_3}}\,\Xi_3,\\  
\langle\tilde{\mathbf{J}}^+_{s_1} \tilde{\mathbf{J}}^-_{s_2} \tilde{\mathbf{J}}^+_{s_3}\rangle&=-\frac{\xi_{1-}\xi_{3-} }{256}\,\frac{\langle J^+_{k_1+\frac{1}{2}}J^-_{k_2+\frac{1}{2}}J^+_{k_3+1}\rangle E}{\langle\overline{2}\,\overline{3}\rangle\sqrt{p_3}}\,\Xi_3,\\
\langle\tilde{\mathbf{J}}^-_{s_1} \tilde{\mathbf{J}}^+_{s_2} \tilde{\mathbf{J}}^+_{s_3}\rangle&=-\frac{\xi_{2-}\xi_{3-} }{256}\frac{\langle J^-_{k_1+\frac{1}{2}}J^+_{k_2+\frac{1}{2}}J^+_{k_3+1}\rangle E}{\langle\overline{3}\,\overline{1}\rangle\sqrt{p_3}}\,\,\Xi_3,
    \end{split}\:
    \begin{split}
\langle\tilde{\mathbf{J}}^+_{s_1} \tilde{\mathbf{J}}^+_{s_2} \tilde{\mathbf{J}}^+_{s_3}\rangle&=\frac{\xi_{1-}\xi_{2-}\xi_{3-} }{2048}\frac{\langle J^+_{k_1+\frac{1}{2}}J^+_{k_2+\frac{1}{2}}J^+_{k_3+1}\rangle E}{\langle12\rangle\sqrt{p_3}}\,\,\Gamma_3,\\
\langle\tilde{\mathbf{J}}^-_{s_1} \tilde{\mathbf{J}}^-_{s_2} \tilde{\mathbf{J}}^+_{s_3}\rangle&=-\frac{\xi_{3-} }{256}\,\frac{\langle J^-_{k_1+\frac{1}{2}}J^-_{k_2+\frac{1}{2}}J^+_{k_3+1}\rangle}{\sqrt{p_3}}\,\Gamma_3,\\
\langle\tilde{\mathbf{J}}^-_{s_1} \tilde{\mathbf{J}}^+_{s_2} \tilde{\mathbf{J}}^-_{s_3}\rangle&=-\frac{\xi_{2-} }{256}\,\frac{\langle J^-_{k_1+\frac{1}{2}}J^+_{k_2+\frac{1}{2}}J^-_{k_3+1}\rangle E}{\langle23\rangle\sqrt{p_3}}\,\Gamma_3,\\
\langle\tilde{\mathbf{J}}^+_{s_1} \tilde{\mathbf{J}}^-_{s_2} \tilde{\mathbf{J}}^-_{s_3}\rangle&=\frac{\xi_{1-} }{256}\,\frac{\langle J^+_{k_1+\frac{1}{2}}J^-_{k_2+\frac{1}{2}}J^-_{k_3+1}\rangle E}{\langle31\rangle\sqrt{p_3}}\,\Gamma_3.
    \end{split}
\end{equation}
This class of correlators are all non homogeneous, that is, their components appear as a sum of bosonic and fermionic structures. We provide a detailed example of one such correlator in appendix \ref{sec:example}.
 \subsection{Double-copy relations}
 Double-copy relations have played an important role in (supersymmetric) scattering amplitudes in the recent decades \cite{Adamo:2022dcm}. Such structures in the non-supersymmetric case were later discovered in conformal field theory \cite{Farrow:2018yni,Lipstein:2019mpu,Jain:2021qcl}. It is thus natural to attempt to extend such results to the supersymmetric case as well. Indeed, in the $\xi_\pm$ variables that we have been using in this paper, such a structure becomes apparent. Let us illustrate this with an example: Consider a three point super-correlator with $s_1=s_2=s_3=2$, that is, the $\langle \mathbf{\tilde{T}}\mathbf{\tilde{T}}\mathbf{\tilde{T}}\rangle$ correlator. As we discussed below equation \eqref{threepointInt}, this correlator is purely homogeneous as it satisfies the triangle inequality. Therefore, it suffices to consider its $(- - -)$ and $(+ + +)$ components. The $(- - -)$ component correlator is,
 \begin{align}\label{TTTmmm}
     \langle \tilde{\mathbf{T}}^-\tilde{\mathbf{T}}^-\tilde{\mathbf{T}}^-\rangle&=\frac{1}{64}\langle T^-T^-T^-\rangle\,\Gamma_3.
    %\\
   % &=c_{2}\frac{1}{64} \frac{\langle12\rangle^2\langle23\rangle^2\langle31\rangle^2 p_1 p_2p_3}{E^6}\Gamma_{3}.
 \end{align}
 Consider now,
 \begin{align}\label{J4J4J4mmm}
      \langle \tilde{\mathbf{J}_4}^-\tilde{\mathbf{J}_4}^-\tilde{\mathbf{J}_4}^-\rangle&=\frac{1}{64}\langle J_4^-J_4^-J_4^-\rangle\,\Gamma_3.
      %\\
      %&=c_{4}\frac{1}{64} \frac{\langle12\rangle^4\langle23\rangle^4\langle31\rangle^4 p_1^3 p_2^3p_3^3}{E^{12}}\Gamma_{3}.
 \end{align}
 If we naively square \eqref{TTTmmm} expecting to obtain \eqref{J4J4J4mmm}, we will obtain zero as $\Gamma_3^2=0$ (evident from the definition \eqref{threepointbuildingblocks}). We note, however, that the relation at the level of \textit{component} correlators,
 \begin{align}
     \langle J_4^{-} J_4^{-} J_4^{-}\rangle\propto p_1 p_2 p_3 \langle T^{-} T^{-} T^{-}\rangle^2,
 \end{align}
 implies at the superfield level \eqref{J4J4J4mmm} that,
 \begin{align}\label{temp1}
      \langle \tilde{\mathbf{J}_4}^-\tilde{\mathbf{J}_4}^-\tilde{\mathbf{J}_4}^-\rangle\propto p_1 p_2 p_3\langle T^{-}T^{-}T^{-}\rangle^2 \Gamma_3.
 \end{align}
Equations \eqref{J4J4J4mmm} and \eqref{temp1} implies a double copy relation between $\langle \mathbf{\tilde{T}}^-\mathbf{\tilde{T}}^-\mathbf{\tilde{T}}^- \rangle$ and $\langle \mathbf{\tilde{J}}_4^-\mathbf{\tilde{J}}_4^-\mathbf{\tilde{J}}_4^- \rangle$. This double copy relation can be extended for arbitrary spin as follows
 \begin{align}
      \langle \tilde{\mathbf{J}}_{s_1}^{-}\tilde{\mathbf{J}}_{s_2}^{-}\tilde{\mathbf{J}}_{s_3}^{-}\rangle&\propto p_1 p_2 p_3\langle J_{s_1'}^{-} J_{s_2'}^{-} J_{s_3'}^{-}\rangle\langle J_{s_1''}^{-} J_{s_2''}^{-} J_{s_3''}^{-}\rangle \Gamma_3,\\
       \langle \tilde{\mathbf{J}}_{s_1}^{+}\tilde{\mathbf{J}}_{s_2}^{+}\tilde{\mathbf{J}}_{s_3}^{+}\rangle&\propto p_1 p_2 p_3\langle J_{s_1'}^{+} J_{s_2'}^{+} J_{s_3'}^{+}\rangle\langle J_{s_1''}^{+} J_{s_2''}^{+} J_{s_3''}^{+}\rangle \Xi_3,
 \end{align}
 where $s_1'+s_1''=s_1,\, s_2'+s_2''=s_2 $, and $s_3'+s_3''=s_3$. Let us also note the spins $s_1, s_2, s_3 \in \mathbb{Z}_{>0}$ and obeys triangle inequality.  %Therefore, we see that the appropriate prescription for double copy at the level of super-correlators is not the naive ``super correlator square" but rather,
% \begin{align}
%     &\langle \tilde{\mathbf{J}}_4^-\tilde{\mathbf{J}}_4^-\tilde{\mathbf{J}}_4^-\rangle\propto p_1 p_2 p_3\langle T^- T^- T^-\rangle \langle \tilde{\mathbf{T}}^-\tilde{\mathbf{T}}^-\tilde{\mathbf{T}}^-\rangle,
 %\end{align}
 %that is, we multiply a superfield correlator with its zeroth component to obtain a higher spin superfield correlator. For general spins $s_1,s_2,s_3\in\mathbb{Z}_{>0}$ obeying the triangle inequality we obtain the universal formula,
 We shall later see in subsection \ref{sec:Neq1toNeq2}, an interesting double copy relation between $\mathcal{N}=1$ and $\mathcal{N}=2$ super correlators.
 \subsection{The super Ward-Takahashi identity}
In this subsection we present the super Ward-Takahashi identity for our non homogeneous correlators. We also choose to work with $\theta$ variables as they suffice for our purposes. The main aim of this subsection is to show that their is a stark distinction between the Grassmann even and Grassmann odd cases.\\ 
 Consider a Grassmann odd correlator $\langle\mathbf{J}_{\frac{3}{2}} \mathbf{J}_{0} \mathbf{J}_{0}\rangle$. The corresponding super Ward-Takahashi identity that we obtain is given by:
\begin{align}
D_{1a}\langle\mathbf{J}_{\frac{3}{2}} ^{abc}\mathbf{J}_{0} \mathbf{J}_{0}\rangle= a_1\;\theta_{1}^{2}\bigg(\slashed p_{2}^{bc}\langle\mathbf{J}_{0}(p_2) \mathbf{J}_{0}(-p_2)\rangle+\slashed p_{3}^{bc}\langle\mathbf{J}_{0}(-p_3) \mathbf{J}_{0}(p_3)\rangle\bigg).
\end{align}
If we consider a Grassmann even correlator such as $\langle\mathbf{T}\mathbf{J}_{0} \mathbf{J}_{0}\rangle$, we find the following super Ward-Takahashi identity:
 \begin{equation}
D_{1a}\langle\mathbf{T}^{abcd}\mathbf{J}_{0}\mathbf{J}_{0}\rangle= a_2\;\theta_{1}^{2}\bigg(\slashed p_{2}^{bc}\,D_{2}^{d}\langle\mathbf{J}_{0}(p_2) \mathbf{J}_{0}(-p_2)\rangle+\slashed p_{3}^{bc}\,D_{3}^{d}\langle\mathbf{J}_{0}(-p_3) \mathbf{J}_{0}(p_3)\rangle\bigg),
 \end{equation}
 where $a_1$ and $a_2$ are proportional to the normalization of the $\langle \mathbf{J}_0\mathbf{J}_0\rangle$ two point super-correlator and the action of covariant derivative is given below equation \eqref{conservation}.
 \subsection{Summary of super correlators: homogeneous and non homogeneous sectors}
In this subsection, we summarize our findings of the structures of the super correlators in the language of homogeneity.
For correlators that obey the triangle inequality, super correlators with total integer spin are homogeneous (with there existing both parity even and odd structures) while those with total half-integer spin are non homogeneous\footnote{In the component language, super correlators that are (non) homogeneous also have component correlators that are also (non) homogeneous.}.
\begin{table}[h!]
    \centering
    \begin{tabular}{|c|c|c|c|}
    \hline
         Total spin& Homogeneous Even &Homogeneous Odd& Non Homogeneous  \\
         \hline
         $s_1+s_2+s_3\in \mathbb{Z}_{>0}$& \checkmark &\checkmark & -\\
         \hline
         $s_1+s_2+s_3\in \mathbb{Z}_{>0}+\frac{1}{2}$& - & -& \checkmark\\ \hline  
    \end{tabular}
    \caption{Correlation functions obeying the triangle inequality $s_i+s_j\ge s_k~\forall i,j,k$.}
    \label{table:table1}
\end{table}\\
For correlators that are not inside the triangle, we have the following results: For integer spin correlators, the individual component correlators appear as differences between the bosonic and fermionic results: $\langle~\rangle_{B}-\langle~\rangle_{F}$. In contrast, for half-integer correlators, they appear as a sum $\langle~\rangle_{B}+\langle~\rangle_{F}$. Further, there is no parity odd structure compatible with the supercurrent conservation in either case.
\begin{table}[h!]
    \centering
    \begin{tabular}{|c|c|c|c|}
    \hline
         Total spin& Boson-Fermion & Boson+Fermion & Parity Odd \\
         \hline
         $s_1+s_2+s_3\in \mathbb{Z}_{>0}$& \checkmark & -& -\\
         \hline
         $s_1+s_2+s_3\in \mathbb{Z}_{>0}+\frac{1}{2}$& - & \checkmark & -\\ \hline  
    \end{tabular}
    \caption{Correlation functions that violate the triangle inequality $s_i+s_j\le s_k$ for some $i,j,k$.}
    \label{table:table2}
\end{table}\\
These results are also consistent with the free theory result. The superfield \eqref{Jsintheta} for the free theory can be written as follows: For integer spins $s$, and half integer spins $k+\frac{1}{2}$ we have respectively\footnote{This can easily be seen from the expressions given for the free theory super currents in \cite{Nizami:2013tpa}.} ($B$ stands for the free bosonic theory result and $F$ stands for the free fermionic theory currents),
\begin{align}
    \mathbf{J}_s&=\big((J_s)_{B}-(J_s)_{F}\big)+\theta\cdot J_{s+\frac{1}{2}}+\partial \big((J_s)_{B}-(J_s)_{F}\big),\notag\\
    \mathbf{J}_{k+\frac{1}{2}}&=J_{k+\frac{1}{2}}+\theta\cdot\big((J_{k})_{B}+(J_{k})_{F}\big)+\partial J_{k+\frac{1}{2}}.
\end{align}
For instance a correlator of three integer spin super currents at the zeroth order (denoted by an $|$ ) is given by,
\begin{align}
    \langle \mathbf{J}_{s_1}\mathbf{J}_{s_2}\mathbf{J}_{s_3}\rangle|=\langle J_{s_1}J_{s_2}J_{s_3}\rangle_{B}-\langle J_{s_1}J_{s_3}J_{s_4}\rangle_{F},
\end{align}
which is the definition of the homogeneous piece. similarly, at all orders in the Grassmann spinor expansion we obtain homogeneous component correlators and hence this is consistent with Table \ref{table:table1}.\\
Similarly, if we consider three half integer spin supercurrents that obey the triangle inequality, we obtain at say $\order{\theta_1\theta_2\theta_3}$ the following expression:
\begin{align}
    \langle \mathbf{J}_{k_1+\frac{1}{2}}\mathbf{J}_{k_2+\frac{1}{2}}\mathbf{J}_{k_3+\frac{1}{2}}\rangle|_{\theta_1\theta_2\theta_3}=\theta_1\theta_2\theta_3\big(\langle J_{k_1+1}J_{k_2+1}J_{k_3+1}\rangle_{B}+\langle J_{k_1+1}J_{k_2+1}J_{k_3+1}\rangle_{F}\big),
\end{align}
which is the definition of the non homogeneous piece and is consistent with\footnote{Similarly, all component correlators in the grassmann spinor expansion are non-homogeneous.} Table \ref{table:table2}.
Repeating the same analysis for the other cases (two integer, one half integer or one integer, two half integer insertions inside and outside the triangle) we find results consistent with Tables \ref{table:table1} and \ref{table:table2}. As for the parity odd solution, which is necessarily homogeneous \cite{Jain:2021whr}, one expects to find one whenever there exists a parity even homogeneous solution as there is little distinction between the two in spinor helicity variables.
\section{Extended supersymmetry: The $\mathcal{N}=2$ case}\label{sec:Neq2}
\subsection{The superspace formulation}
The $\mathcal{N}=2$ superspace can be described by the coordinates $(x^\mu,\theta^a,\Bar{\theta}^a)$ where $x^\mu$ are the usual position space coordinates and $\theta^a$ and $\Bar{\theta}^a$ are two component Grassmann spinors \cite{Park:1999cw,Jain:2022izp}. The generators of the $\mathcal{N}=2$ superconformal algebra include the usual conformal generators $P_\mu,M_{\mu\nu},D,K_\mu$ along with the supersymmetry generators $Q_a,\Bar{Q}_a$ and the special superconformal generators $S_a$ and $\Bar{S}_a$. More information on this formalism as well as the action of the generators on primary superfields can be found for instance in \cite{Nizami:2013tpa,Jain:2022izp}.

We now, parallel our analysis of the $\mathcal{N}=1$ case in section \ref{sec:Neq1Formalism}, first convert $x^\mu$ to $p_\mu$ by a Fourier transform. We then express $p_\mu$ in spinor helicity variables as in \eqref{pmuspinorhelicity}. The key change that arises in the $\mathcal{N}=2$ is the fact the $\theta^a$ is no longer Hermitian. Therefore, it has to be described by two complex Grassmann variables in contrast to the single complex Grassmann variables $\eta$ \eqref{etavariables}. The $\mathcal{N}=2$ Grassmann spinors can be expressed as follows: For complex Grassmann variables $\eta,\mu$, we define,
\begin{align}\label{etamu}
    &\theta^a=\frac{\Bar{\eta}\lambda^a+\mu \Bar{\lambda}^a}{2p},\notag\\
    &\Bar{\theta}^a=\frac{\eta\Bar{\lambda}^a+\Bar{\mu}\lambda^a}{2p}.
\end{align}
Further, just as we performed a Grassmann twistor transform in the $\mathcal{N}=1$ case \eqref{GrassmannTwistorTransform}, we do so in the $\mathcal{N}=2$ case as follows. We define,
\begin{align}\label{GTTNeq2}
    \Tilde{F}(\eta,\chi,\mu,\nu)=\int d\Bar{\eta}~d\Bar{\mu}~e^{-\frac{\chi\Bar{\eta}}{4}-\frac{\nu \Bar{\mu}}{4}}F(\eta,\Bar{\eta},\mu,\Bar{\mu}).
\end{align}
We then define the coordinates,
\begin{align}\label{xiomegavariables}
&\xi_{\pm}=\chi\pm\eta,\quad\omega_{\pm}=\nu\pm \mu.
\end{align}
In terms of these coordinates, the supersymmetry generators of the $\mathcal{N}=2$ superconformal algebra take the following simple forms:
\begin{equation}\label{xiomegaspacegen}
    Q_a=2\lambda_a\frac{\partial}{\partial\omega_+}+\frac{\Bar{\lambda}_a}{4}\xi_+,\qquad
    \Bar{Q}_a=2\lambda_a\frac{\partial}{\partial\xi_+}+\frac{\Bar{\lambda}_a}{4}\omega_+.
\end{equation}
The remaining generators are identical to their $\mathcal{N}=1$ counterparts given in \eqref{xispacegen}. However, there is one very important distinction between the $\mathcal{N}=1$ and $\mathcal{N}=2$ cases, that is, the presence of a $U(1)$ R-symmetry. The $Q$ and $\Bar{Q}$ operators defined in \eqref{xiomegaspacegen} have $R$ charges $-1$ and $+1$ respectively. This implies the following commutation relations
\begin{equation}\label{Rcommutators}
  [R,Q_a]=-Q_a,\quad 
    [R,\Bar{Q}_a]=+\Bar{Q}_a.
\end{equation}
The representation of $R$ in the $\xi_{\pm},\omega_{\pm}$,  acting on primary super fields is the following:
\begin{align}\label{Roperator}
    R=\omega_+\frac{\partial}{\partial\omega_+}-\xi_+\frac{\partial}{\partial \xi_+},
\end{align}
where $R$ is the symmetry generator. It can easily be checked using \eqref{xiomegaspacegen} and \eqref{Roperator} that the commutators \eqref{Rcommutators} hold. 
\subsection{Super-currents and Ward identities}
We are interested in constraining correlators involving symmetric traceless conserved super-currents. The general form of such currents in  $\mathcal{N}=2$ SCFTs can be found for instance in \cite{Nizami:2013tpa}. Here, we present its momentum superspace avatar:
\begin{align}\label{Neq2Js}
\mathbf{J}^{a_1 a_2 ... a_{2s}}_{s}&=J^{a_1 a_2 ... a_{2s}}_{s}+\theta_m J^{a_1 a_2 ... a_{2s}m}_{s+\frac{1}{2}}+\overline{\theta}_m \overline{J}^{a_1 a_2 ... a_{2s}m}_{s+\frac{1}{2}}+\theta_m\overline{\theta}_nJ^{a_1 a_2 ... a_{2s}mn}_{s+1}+\frac{\theta_m\overline{\theta}^m}{2}\slashed{p}^{a_1}_n J^{a_2 ... a_{2s}}_{s}\notag\\&
-\frac{\theta^2 \overline{\theta}_m}{4}\slashed{p}^m_n J^{a_1 a_2 ... a_{2s}n}_{s+\frac{1}{2}}-\frac{\overline{\theta}^2 \theta_m}{4}\slashed{p}^m_n \overline{J}^{a_1 a_2 ... a_{2s}n}_{s+\frac{1}{2}}-\frac{\theta^2\overline{\theta}^2 }{16}p^2 J^{a_1 a_2 ... a_{2s}}_{s}.
\end{align}
We then use the definition \eqref{etamu} and perform the Grassmann twistor transform \eqref{GTTNeq2}.
Their is a drastic simplification compared to \eqref{Neq2Js} when the result is expressed in the $(\xi_\pm,\omega_\pm)$ variables \eqref{xiomegavariables}. We have,
\begin{align}\label{Jsinxiomega}
    &\mathbf{\tilde{J}_s^-}=\frac{1}{4}\bigg[\frac{\xi_+\omega_+}{4}J_s^--\frac{1}{2\sqrt{p}}\bigg(\omega_+J_{s+\frac{1}{2}}^--\xi_+\Bar{J}_{s+\frac{1}{2}}^{-}\bigg)-\frac{J_{s+1}^-}{p}\bigg],\notag\\
    &\mathbf{\tilde{J}_s^+}=\frac{\xi_-\omega_-}{16}\bigg[J_s^++\frac{1}{4\sqrt{p}}\bigg(\omega_+ J_{s+\frac{1}{2}}^++\xi_+\Bar{J}_{s+\frac{1}{2}}^+\bigg)-\frac{\xi_+\omega_+}{16 p}J_{s+1}^+\bigg].
\end{align}
    We can then construct super-correlators using \eqref{Jsinxiomega}, as we did for the $\mathcal{N}=1$ case. The component correlators can then be constrained using the action of $Q_a$ and $\Bar{Q}_a$, as given in \eqref{xiomegaspacegen}. The $R$ symmetry is going to play an important role here as correlators have to be $R$ symmetric. Before proceeding further, we note that, $\xi_{i+}$ has $R$ charge $-1$, while $\omega_{i+}$ has $R$ charge $+1$. Then, correlators are forced to be made of elements of form $\xi_{i+}\omega_{i+}$, where $i$ is the operator label. This is exactly what we got and we present our findings below for two and three point functions. 

\subsection{Two point functions}
The $\mathcal{N}=2$ two point functions for any integer spin $s$ take the following forms in the two independent helicity configurations:
\begin{align}\label{Neq2Twopoint}
    \langle \mathbf{\Tilde{J}_s^-}\mathbf{\Tilde{J}_s^-}\rangle&=\frac{\langle 1 2\rangle^{2s}}{16\,p_1}\bigg(\frac{\xi_{1+}\omega_{1+}\xi_{2+}\omega_{2+}}{16}-\frac{\langle 1 2\rangle}{4p_1}(\omega_{1+}\xi_{2+}+\xi_{1+}\omega_{2+})+\frac{\langle 1 2\rangle^2}{p_1^2}\bigg),\notag\\
    \langle \mathbf{\Tilde{J}_s^+}\mathbf{\Tilde{J}_s^+}\rangle&=\xi_{1-}\omega_{1-}\xi_{2-}\omega_{2-}\frac{\langle \Bar{1} \Bar{2}\rangle^{2s+3}}{65536 \,p_1^4}\bigg(\frac{\xi_{1+}\omega_{1+}\xi_{2+}\omega_{2+}}{16}-\frac{\langle 1 2\rangle}{4p_1}(\omega_{1+}\xi_{2+}+\xi_{1+}\omega_{2+})+\frac{\langle 1 2\rangle^2}{p_1^2}\bigg).
\end{align}
Further, the building block of  $\mathcal{N}=2$ two-point functions can be written as a product of two individual building blocks.
Then, the correlators can be written as,
\begin{align}\label{producttwo}
    \langle \mathbf{\Tilde{J}_s^-}\mathbf{\Tilde{J}_s^-}\rangle&=\frac{\langle 1 2\rangle^{2s}}{16\,p_1}\bigg(\frac{\xi_{1+}\omega_{1+}}{4}-\frac{\langle 1 2\rangle}{p_1}\bigg)\bigg(\frac{\xi_{2+}\omega_{2+}}{4}-\frac{\langle 1 2\rangle}{p_1}\bigg),\notag\\
    \langle \mathbf{\Tilde{J}_s^+}\mathbf{\Tilde{J}_s^+}\rangle&=\xi_{1-}\omega_{1-}\xi_{2-}\omega_{2-}\frac{\langle \Bar{1} \Bar{2}\rangle^{2s+3}}{65536 \,p_1^4}\bigg(\frac{\xi_{1+}\omega_{1+}}{4}-\frac{\langle 1 2\rangle}{p_1}\bigg)\bigg(\frac{\xi_{2+}\omega_{2+}}{4}-\frac{\langle 1 2\rangle}{p_1}\bigg).
\end{align}
It can be clearly seen after the discussion in the above section, that the correlators are formed by the building blocks which are inherently R symmetric as they are made of elements like $\xi_{i+}\omega_{i+}$. 
%This constrain of being R symmetric helps us to establish super double copy relations as it will be dis subsection \ref{sec:Neq1toNeq2}.%

\subsection{Three point function}
The structure of the three point function in the $\mathcal{N}=2$ theories are much more complicated than their $\mathcal{N}=1$ counterpart. However, by a careful analysis, we were able still able to obtain the three point correlators in the $\mathcal{N}=2$ theories for any arbitrary integer spin in terms of the following building blocks which is reminiscent of the $\mathcal{N}=1$ case \eqref{threepointbuildingblocks}.
\begin{align}\label{Neq2buildingblocksthreepoint}
   \nonumber \Omega_1&=\Big(\langle23\rangle^2 \xi_{1+}\omega_{1+}+\langle23\rangle\langle31\rangle\xi_{1+}\omega_{2+}+\langle12\rangle\langle23\rangle \xi_{1+}\omega_{3+}+\langle23\rangle\langle31\rangle\xi_{2+}\omega_{1+}+\langle31\rangle^2 \xi_{2+}\omega_{2+}\\
\nonumber&\qquad+\langle12\rangle\langle31\rangle \xi_{2+}\omega_{3+}+\langle12\rangle\langle23\rangle\xi_{3+}\omega_{1+}+\langle12\rangle\langle31\rangle \xi_{3+}\omega_{2+}+\langle12\rangle^2\xi_{3+}\omega_{3+}\Big),\\
\nonumber\Omega_2&= \Big(-\langle23\rangle \xi_{1+}\omega_{1+} \big(\xi_{2+}\omega_{3+}-\xi_{3+}\omega_{2+}\big)+\langle 31\rangle \xi_{2+}\omega_{2+}\big(\xi_{1+}\omega_{3+}-\xi_{3+}\omega_{1+}\big)\\
&\qquad-\langle 1 2\rangle \xi_{3+}\omega_{3+}\big(\xi_{1+}\omega_{2+}-\xi_{2+}\omega_{1+}\big)\Big),\notag\\
\Omega_3&=\xi_{1+}\xi_{2+}\xi_{3+}\omega_{1+}\omega_{2+}\omega_{3+}.
\end{align}
For example, in the $(- - -)$ helicity we obtain\footnote{We actually obtain a two parameter family of solutions. However, one of them is inconsistent with the operator product expansion in the sense of \cite{Jain:2021whr} and hence we set it to zero. More details on the same and more will appear in a subsequent paper \cite{jain2024scft}.},
\begin{align}\label{Neq2ThreePointmmm}
    &\langle \mathbf{\tilde{J}_{s_1}}^-\mathbf{\tilde{J}_{s_2}}^-\mathbf{\tilde{J}_{s_3}}^-\rangle=\langle J_{s_1}^{-}J_{s_2}^{-}J_{s_3}^{-}\rangle\bigg(\frac{\Omega_1}{E^2}+\frac{\Omega_2}{8 E}-\frac{\Omega_3}{64}\bigg).
\end{align}
For spins that satisfy the triangle inequality, \eqref{Neq2ThreePointmmm} is homogeneous similar to it's $\mathcal{N}=1$ \eqref{threepointInt}.
It is straightforward to obtain analogous formulae in the remaining helicity configurations as well.
\subsection{Double copy: $\mathcal{N}=1\otimes \mathcal{N}=1\to\mathcal{N}=2$}\label{sec:Neq1toNeq2}
\subsection*{Two-Point function}
Consider the following integer two point functions  in the $\mathcal{N}=1$ theory in the $(- -)$ helicity configuration \eqref{twopointintspin},
\begin{align}\label{twopointintspinfordoublecopy}
    &\langle \mathbf{\Tilde{J}}_s^-(\xi_{1+})\mathbf{\Tilde{J}}^-_s(\omega_{2+})\rangle_{\mathcal{N}=1}=\frac{\langle 1 2\rangle^{2s}}{16 p_1}\bigg( \xi_{1+}\omega_{2+}-\frac{4\langle 1 2\rangle}{p_1}\bigg),\notag\\
    &\langle \mathbf{\Tilde{J}}^-_s(\omega_{1+})\mathbf{\Tilde{J}}^-_s(\xi_{2+})\rangle_{\mathcal{N}=1}=\frac{\langle 1 2\rangle^{2s}}{16 p_1}\bigg( \omega_{1+}\xi_{2+}-\frac{4\langle 1 2\rangle}{p_1}\bigg),
\end{align}
where we have made the dependence on the Grassmann twistor variables explicit. The reason for the strange choice of the Grassmann twistor variable dependence of the super currents in both lines of \eqref{twopointintspinfordoublecopy} will become clear shortly.
We now take a product between the first and second line of \eqref{twopointintspinfordoublecopy}. The result is,
\small
\begin{align}
    \langle \mathbf{\Tilde{J}}^-_s(\xi_{1+})\mathbf{\Tilde{J}}^-_s(\omega_{2+})\rangle_{\mathcal{N}=1}\langle \mathbf{\Tilde{J}}^-_s(\omega_{1+})\mathbf{\Tilde{J}}^-_s(\xi_{2+})\rangle_{\mathcal{N}=1}&=\frac{\langle 1 2\rangle^{4s}}{16 p_1^2}\bigg(\frac{\xi_{1+}\omega_{1+}\xi_{2+}\omega_{2+}}{16}-\frac{\langle 1 2\rangle}{4p_1}(\omega_{1+}\xi_{2+}+\xi_{1+}\omega_{2+})+\frac{\langle 1 2\rangle^2}{p_1}\bigg).
\end{align}
\normalsize
By comparing the RHS of the above equation with the $\mathcal{N}=2$ two point function \eqref{Neq2Twopoint}, we obtain a super double copy relation!
\begin{align}\label{Neq1toNeq2DoubleCopy}
    \langle \mathbf{\Tilde{J}}^-_s(\xi_{1+})\mathbf{\Tilde{J}}^-_s(\omega_{2+})\rangle_{\mathcal{N}=1}\langle \mathbf{\Tilde{J}}^-_s(\omega_{1+})\mathbf{\Tilde{J}}^-_s(\xi_{2+})\rangle_{\mathcal{N}=1}=\frac{1}{p_1} \langle \mathbf{\Tilde{J}}_{2s}^{-}(\xi_{1+},\omega_{1+}) \mathbf{\Tilde{J}}_{2s}^{-}(\xi_{2+},\omega_{2+})\rangle_{\mathcal{N}=2}.
\end{align}
The reason for the weird choices of Grassmann twistor variable dependence in \eqref{twopointintspinfordoublecopy} is now clear. This and only this particular choice will ensure that the RHS of \eqref{Neq1toNeq2DoubleCopy}, which is a $\mathcal{N}=2$ two-point function will be invariant under the action of the $R$ symmetry generator \eqref{Roperator}. Similar results in the other helicity configuration as well as double copy relations involving half integer $\mathcal{N}=1$ supercorrelators can also easily be obtained, all thanks to the simplicity of the Grassmann twistor variables.
\subsection*{Three-Point function}
Let us now proceed to the three point case. For simplicity, we consider the cases that satisfy the triangle inequality $s_i+s_j\ge s_k \,\forall i,j,k\in{1,2,3}$. Our prescription for obtaining $\mathcal{N}=2$ super correlators from products of $\mathcal{N}=1$ super correlators is constructing the product such that the resulting object possesses the required $R$ symmetry property. Our goal is to take products of $\mathcal{N}=1$ three-point functions \eqref{threepointInt} to reproduce the $\mathcal{N}=2$ answer \eqref{Neq2ThreePointmmm}. From the required $U(1)$ R symmetry of the result, we know that the total number of $\xi_{i+}$ and $\omega_{i+}$ variables appearing in each term must be equal. This then restricts our ansatz for the $\mathcal{N}=1$ product.
Let us concentrate on the $(- - -)$ helicity configuration\footnote{One can carry out an analogous analysis in the other helicities.}. We take,
\begin{align}\label{AnsatzForDoubleCopy}
    \textbf{Ansatz}=&\bigg[a_1\langle \mathbf{\Tilde{J}}_s^{-}(\xi_{1+})\mathbf{\Tilde{J}}_s^{-}(\xi_{2+})\mathbf{\Tilde{J}}_s^{-}(\xi_{3+})\rangle_{\mathcal{N}=1} \langle \mathbf{\Tilde{J}}_s^{-}(\omega_{1+})\mathbf{\Tilde{J}}_s^{-}(\omega_{2+})\mathbf{\Tilde{J}}_s^{-}(\omega_{3+})\rangle_{\mathcal{N}=1}\notag\\
    &+a_2\langle \mathbf{\Tilde{J}}_s^{-}(\xi_{1+})\mathbf{\Tilde{J}}_s^{-}(\xi_{2+})\mathbf{\Tilde{J}}_s^{-}(\omega_{3+})\rangle_{\mathcal{N}=1} \langle \mathbf{\Tilde{J}}_s^{-}(\xi_{1+})\mathbf{\Tilde{J}}_s^{-}(\xi_{2+})\mathbf{\Tilde{J}}_s^{-}(\omega_{3+})\rangle_{\mathcal{N}=1}\notag\\
    &+a_3\langle \mathbf{\Tilde{J}}_s^{-}(\xi_{1+})\mathbf{\Tilde{J}}_s^{-}(\omega_{2+})\mathbf{\Tilde{J}}_s^{-}(\xi_{3+})\rangle_{\mathcal{N}=1} \langle \mathbf{\Tilde{J}}_s^{-}(\omega_{1+})\mathbf{\Tilde{J}}_s^{-}(\xi_{2+})\mathbf{\Tilde{J}}_s^{-}(\omega_{3+})\rangle_{\mathcal{N}=1}\notag\\
    &+a_4\langle \mathbf{\Tilde{J}}_s^{-}(\omega_{1+})\mathbf{\Tilde{J}}_s^{-}(\xi_{2+})\mathbf{\Tilde{J}}_s^{-}(\xi_{3+})\rangle_{\mathcal{N}=1} \langle \mathbf{\Tilde{J}}_s^{-}(\xi_{1+})\mathbf{\Tilde{J}}_s^{-}(\omega_{2+})\mathbf{\Tilde{J}}_s^{-}(\omega_{3+})\rangle_{\mathcal{N}=1}\bigg].
\end{align}
We then demand that the ansatz \eqref{AnsatzForDoubleCopy} is invariant under the simultaneous action of the $R$ symmetry generator \eqref{Roperator} on all the insertions. This yields the constraint $a_1=a_2=a_3=a_4$. In fact \eqref{AnsatzForDoubleCopy} becomes,
\begin{align}
    &\langle J_{s}^{-} J_{s}^{-}J_{s}^{-}\rangle^2 \bigg(\frac{\Omega_1}{E^2}+\frac{\Omega_2}{8 E}-\frac{\Omega_3}{64}\bigg),
\end{align}
where the $\Omega_i$ are the $\mathcal{N}=2$ building blocks \eqref{Neq2buildingblocksthreepoint}. Therefore, we obtain a super double copy,
\begin{align}\label{mainDoubleCopy}
    &\langle \mathbf{\Tilde{J}}_{2s_1}^{-}(\xi_{1+},\omega_{1+})\mathbf{\Tilde{J}}_{2s_2}^{-}(\xi_{2+},\omega_{2+})\mathbf{\Tilde{J}}_{2s_3}^{-}(\xi_{3+},\omega_{3+})\rangle_{\mathcal{N}=2}\notag\\
    &\quad=p_1p_2p_3\Big[\langle \mathbf{\Tilde{J}}_s^{-}(\xi_{1+})\mathbf{\Tilde{J}}_s^{-}(\xi_{2+})\mathbf{\Tilde{J}}_s^{-}(\xi_{3+})\rangle_{\mathcal{N}=1} \langle \mathbf{\Tilde{J}}_s^{-}(\omega_{1+})\mathbf{\Tilde{J}}_s^{-}(\omega_{2+})\mathbf{\Tilde{J}}_s^{-}(\omega_{3+})\rangle_{\mathcal{N}=1}\notag\\
    &\qquad~\,\quad\quad+\langle \mathbf{\Tilde{J}}_s^{-}(\xi_{1+})\mathbf{\Tilde{J}}_s^{-}(\xi_{2+})\mathbf{\Tilde{J}}_s^{-}(\omega_{3+})\rangle_{\mathcal{N}=1} \langle \mathbf{\Tilde{J}}_s^{-}(\xi_{1+})\mathbf{\Tilde{J}}_s^{-}(\xi_{2+})\mathbf{\Tilde{J}}_s^{-}(\omega_{3+})\rangle_{\mathcal{N}=1}\notag\\
    &\qquad~\,\quad\quad+\langle \mathbf{\Tilde{J}}_s^{-}(\xi_{1+})\mathbf{\Tilde{J}}_s^{-}(\omega_{2+})\mathbf{\Tilde{J}}_s^{-}(\xi_{3+})\rangle_{\mathcal{N}=1} \langle \mathbf{\Tilde{J}}_s^{-}(\omega_{1+})\mathbf{\Tilde{J}}_s^{-}(\xi_{2+})\mathbf{\Tilde{J}}_s^{-}(\omega_{3+})\rangle_{\mathcal{N}=1}\notag\\
    &\qquad~\,\quad\quad+\langle \mathbf{\Tilde{J}}_s^{-}(\omega_{1+})\mathbf{\Tilde{J}}_s^{-}(\xi_{2+})\mathbf{\Tilde{J}}_s^{-}(\xi_{3+})\rangle_{\mathcal{N}=1} \langle \mathbf{\Tilde{J}}_s^{-}(\xi_{1+})\mathbf{\Tilde{J}}_s^{-}(\omega_{2+})\mathbf{\Tilde{J}}_s^{-}(\omega_{3+})\rangle_{\mathcal{N}=1}\Big].
\end{align}
Notice that the result \eqref{mainDoubleCopy} is true for any integer spins $s_1,s_2$ and $s_3$ that obey the triangle inequality. The double copy at the level of homogeneous component correlators was essential for this result. For non homogeneous correlators, there exist some double copy relations albeit more complicated ones. Our prescription to obtain $\mathcal{N}=2$ super-correlators from R symmetry preserving products of $\mathcal{N}=1$ super-correlators could potentially generalize to higher points. We postpone such investigations to an upcoming paper \cite{jain2024scft}.
\section{Discussion and future directions}\label{sec:discussion}
\subsection*{Summary}
In this paper, we have developed a formalism to study three dimensional super conformal field theories. Rather than working in momentum space with the usual grassmann spinors $\theta^a$, we found it advantageous to express both the momenta and the grassmann spinors in spinor helicity variables, i.e, super spinor helicity variables. Looking at the structures that one obtains in these variables motivated us to perform a grassmann twistor transform where we noticed a remarkable simplification at the level of symmetry generators and super-correlation functions. We then went on to constrain and solve for all (half) integer two and three point functions in $\mathcal{N}=1$ SCFTs in these new variables. In contrast to the case by case approach like in the earlier position space analysis, we obtain a universal structure for these super-correlators. This immediately gave us a double copy relation at the level of the super-correlators as well as give us a structure reminiscent of the four dimensional flat space scattering amplitudes. We also developed the formalism for the $\mathcal{N}=2$ case where we discovered similar universal structures for the correlators independent of spin. Further, we presented a prescription to obtain $\mathcal{N}=2$ super correlators from their $\mathcal{N}=1$ counterparts through a super double copy relation.
\subsection*{Future directions}
\subsection*{Extended supersymmetry}
The most natural generalization of our work is to extend our formalism to theories with higher supersymmetry. The spinor helicity variable representation of $\theta_a^A$ where $A$ is the R-symmetry index is a simple generalization of the $\mathcal{N}=1$ definition \eqref{etavariables}.
\begin{align}
    \theta_a^A=\frac{\Bar{\eta}^A \lambda_a+\eta^A\Bar{\lambda}_a}{2p},
\end{align}
which is just obtained from \eqref{etavariables} by promoting the complex grassmann variable $\eta$ to $\eta^A$ to accommodate the R-symmetry. Likewise, one can obtain the generators and the super-currents and perform an analysis parallel to the one presented in this paper to obtain correlation functions in these theories as well.
\subsection*{(Slightly broken) higher spin supersymmetry}
Yet another important extension of our work is the study of higher spin super conformal field theories. It would be interesting to see the interplay between higher spin symmetry and super symmetry as both these symmetries connect correlators with different amounts of spin. Further, the case when the higher spin symmetry is slightly broken is yet another avenue that is worth exploring \cite{Maldacena:2012sf,Maldacena:2011jn,Giombi:2016zwa,Turiaci:2018nua,Aharony:2018npf,Skvortsov:2018uru,Li:2019twz,Kalloor:2019xjb,Silva:2021ece,Binder:2021cif,Gerasimenko:2021sxj}. Supersymmetric Chern Simons+matter theories provide one such quintessential example. We plan to pursue the development of the supersymmetric slightly broken higher spin algebra to analyze and solve for correlation functions in the near future. This will allow us to obtain the supersymmetric analogue of results such as those obtained in \cite{Jain:2022ajd} where all n point functions involving single trace primary operators in CS+fermionic matter theory were obtained in terms of the free theory correlators. The maximally supersymmetric version of the CS+matter theory is the $\mathcal{N}=6$ ABJM theory. We hope that our analysis will prove also prove useful in the pursuit of its study.
\subsection*{A simplex representation for spinning correlators}
A simplex representation for four (and higher) point scalar correlators in CFT has been obtained \cite{Bzowski:2019kwd,Bzowski:2020kfw,Caloro:2022zuy} . However, an extension to the spinning cases has not yet been achieved. Supersymmetry, as it connects correlators with different spins could provide a way to obtain analogous simplex representations for spinning correlators starting with the known expression for scalars. Extending the technology of weight shifting and spin raising operators to supersymmetric theories also is also another venue of interest. 
\subsection*{Making a precise connection to the four dimensional S matrices}
The fascinating connection between non supersymmetric correlators in three dimensions and flat space amplitudes in four dimensions has been analyzed thoroughly and well understood. A similar statement for supersymmetric theories has not yet been obtained. Four dimensional super symmetric scattering amplitudes take extremely simple forms and whilst our building blocks for three point super-correlators \eqref{threepointbuildingblocks} are similar in form to the four dimensional three point amplitude building blocks \cite{Elvang:2011fx,Elvang:2013cua}, we have not attempted to establish a precise connection between the two. Such a connection could facilitate a number of interesting directions such as importing recursion relation techniques such as the BCFW recursion relation to three dimensional SCFTs, color kinematics duality for CFT correlators and so on.
\acknowledgments
We would like to thank A. Nizami, A.Jain, and T. Sharma for valuable discussions and collaboration in the initial stages of this work. We would also like to thank E.Buchbinder and B.Stone for helpful email communication. We also thank M.Ali, N.Bhave and S.Pant for useful discussions. S.Y acknowledges support from INSPIRE fellowship from the Department of Science and Technology, Government of India. We acknowledge our debt to the people of India
for their steady support of research in basic sciences.
\appendix
\section{Notation, Conventions and some useful formulae}\label{sec:notation}
In this appendix, we summarize our conventions and notations as well as provide some formulae that we found useful.
\subsection*{Notations}
Momentum conservation in all our (super) correlation functions is implicit, i.e we do not explicitly write the momentum conserving Dirac-delta functions.
The component fields are presented in italics, whereas superfields are always presented in bold typeface. For instance, $J_s$ is a component field whereas $\mathbf{J}_s$ is a super field.
Superfields when expressed in the grassmann twistor variables are denoted in bold and with a tilde such as $\mathbf{\Tilde{J}}_s$. The subscripts b or B and f or F for component correlators refer to the free bosonic and free fermionic theory correlators respectively.
\subsection*{Conventions}
We work with the usual flat Euclidean metric,
\begin{align}
    \delta_{\mu\nu}=\textbf{diag}(1,1,1),
\end{align}
with which vector indices are raised and lowered. Since upper and lower indices are identical in this case, we do not need to distinguish between them. Our convention for the three dimensional Levi-Civita symbol is,
\begin{align}
\epsilon_{123}&=\epsilon^{123}=1,\notag\\
    \epsilon_{\mu\nu\rho}\epsilon_{\alpha\beta\rho}&=\delta_{\mu\alpha}\delta_{\nu\beta}-\delta_{\mu\beta}\delta_{\nu\alpha}.
\end{align}
Spinor indices on the other hand are raised and lowered using the two dimensional Levi-Civita symbol $\epsilon_{ab}$ which is given by,
\begin{align}
\epsilon_{12}&=\epsilon^{12}=1,\notag\\
    \epsilon_{ab}\epsilon^{ac}&=\delta_b^c.
\end{align}
Our convention for raising and lowering indices is as follows: Given a spinor $A_a$ we have,
\begin{align}\label{IndexRaisingSpinor}
    A^a=\epsilon^{ab}A_b\iff A_a=\epsilon_{ba}A^b.
\end{align}
Further, spinorial derivatives in our conventions are as follows:
\begin{align}
    \frac{\partial A^a}{\partial A^b}=\delta^a_b.
\end{align}
However, in contrast to the index raising and lowering for spinors \eqref{IndexRaisingSpinor}, we have the following conventions for the derivatives:
\begin{align}
    \epsilon^{ab}\frac{\partial}{\partial A^b}=-\frac{\partial}{\partial A_a}.
\end{align}
We choose the following representations of the Pauli matrices 
\begin{align}
    &\sigma^1=\begin{pmatrix}0&1\\
    1&0\end{pmatrix},\sigma^2=\begin{pmatrix}0&-i\\
    i&0\end{pmatrix},\sigma^3=\begin{pmatrix}1&0\\
    0&-1\end{pmatrix},
\end{align}
which satisfy,
\begin{align}
    (\sigma^\mu)^a_b(\sigma^\nu)^c_a=\delta^{\mu\nu}\delta^c_b+i\epsilon^{\mu\nu\rho}(\sigma^\rho)^c_b.
\end{align}
\subsection*{Three dimensional spinor-helicity variables}

Given a three-vector $p_\mu$, we can trade it for a matrix $(\slashed{p})^{a}_{b}$ in the following way:
\begin{align}
    (\slashed{p})^{a}_{b}=p_\mu(\sigma^\mu)^a_b=\lambda_b\Bar{\lambda}^a+p\delta^a_b.
\end{align}
We can extract the magnitude of the momentum (energy) through the following bracket:
\begin{equation}
    p=-\frac{1}{2}\langle \lambda \Bar{\lambda}\rangle.
\end{equation}
We can also contract form spinor dot products belonging to different momenta in the following way:
\begin{align}
    &\langle i j\rangle=\lambda_{ia}\lambda_j^b.
\end{align}
 Since we work with spinning operators, we require the use of polarization vectors. Our conventions for the same are,
\begin{align}
    &z_\mu^{-}(\sigma^\mu)^a_b=(\slashed{z}^{-})^a_b=\frac{\lambda_b\lambda^a}{p},\\
    &z_\mu^{+}(\sigma^\mu)^a_b=(\slashed{z}^{+})^a_b=\frac{\Bar{\lambda}_b\Bar{\lambda}^a}{p}.
\end{align}
%Polarization vectors of higher spin operators can be formed by taking products of these spin one polarization vectors.\\%
For a spin half operator we define the polarization spinor to be,
\begin{align}
    \zeta_a^{-}=\frac{\lambda_a}{\sqrt{p}},\qquad\zeta_a^{+}=\frac{\Bar{\lambda}_a}{\sqrt{p}}.
\end{align}
In a correlation function involving $n$ operator insertions, momentum conservation reads,
\begin{align}
    &\sum_{i=1}^{n}p_i^\mu=0.
\end{align}
Contracting this equation with $(\sigma_\mu)^a_b$ yields momentum conservation in terms of the spinor variables.
\begin{align}\label{momconsSH}
    \sum_{i=1}^{n}\lambda_{ib}\Bar{\lambda}_i^a=-E\delta^a_b.
\end{align}
The three dimensional dot product of two three vectors $x$ and $y$ can be written in spinor notation using,
\begin{align}
    &x\cdot y=\frac{1}{2}(\slashed{x})^a_b(\slashed{y})^b_a.
\end{align}
Since we work with parity odd correlation functions as well, we will require the following formula:
\begin{align}
&\epsilon^{\mu\nu\rho}=\frac{1}{2i}(\sigma^\mu)^a_b(\sigma^\nu)^c_a(\sigma^\rho)^b_c.
\end{align}
For any three vectors $v_1,v_2$ and $v_3$ we define for convenience,
\begin{equation}
    \epsilon^{v_1 v_2 v_3}=v_{1\mu}v_{2\nu}v_{3\rho}\epsilon^{\mu\nu\rho}.
\end{equation}
\subsection*{Some useful spinor-helicity variables identities}
Contracting the momentum conservation equation \eqref{momconsSH} with different combinations of spinors, we get
\begin{align}
    &\langle j i\rangle\langle\Bar{i}\Bar{k}\rangle=E\langle j\Bar{k}\rangle,\\
    &\langle j i\rangle\langle \Bar{i}k\rangle=(E-2p_k)\langle j k\rangle,\\
    &\langle \Bar{j}i\rangle\langle\Bar{i}k\rangle=(E-2p_j)\langle \Bar{j}\Bar{k}\rangle,\\
    &\langle \Bar{j}i\rangle\langle\Bar{i}k\rangle=(E-2p_j-2p_k)\langle \Bar{j}k\rangle,\\
    &\langle j i\rangle\langle\Bar{i}j\rangle+\langle j k\rangle\langle\Bar{k}j\rangle=0,\\
    &\langle i j\rangle\langle\Bar{i}\Bar{j}\rangle=E(E-2p_k),
\end{align}
where $i$, $j$ and $k$ are all distinct labels.\\
The dot products of the polarizations with the momenta are given by,
\begin{equation}
\begin{split}
    p_i\cdot z_j^{-}&=-\frac{\langle i j\rangle\langle\Bar{i}j\rangle}{2 p_j},\\
    p_i\cdot z_j^{+}&=-\frac{\langle i\Bar{j}\langle\Bar{i}\Bar{j}\rangle}{2 p_j},
    \end{split}\qquad
    \begin{split}
    z_i^{-}\cdot z_j^{-}&=-\frac{\langle ij\rangle^2}{2 p_i p_j},\\
    z_i^{-}\cdot z_{j}^{+}&=-\frac{\langle i \Bar{j}\rangle^2}{2 p_i p_j},\\
    z_i^{+}\cdot z_j^{+}&=-\frac{\langle\Bar{i}\Bar{j}\rangle^2}{2 p_i p_j}.
    \end{split}
\end{equation}
For the contractions of momentum and polarization vectors with the three dimensional Levi-Civita symbol we have,
\begin{equation}
\begin{split}
\epsilon^{z_i^{-}z_j^{-}p_k}&=i\frac{\langle i j\rangle\left(\langle i\Bar{k}\rangle\langle k j\rangle+\langle i j\rangle p_k\right)}{2 p_i p_j},\\
\epsilon^{z_i^{-}z_j^{+}p_k}&=i\frac{\langle i \Bar{j}\rangle\left(\langle i\Bar{k}\rangle\langle k \Bar{j}\rangle+\langle i \Bar{j}\rangle p_k\right)}{2 p_i p_j},\\
\epsilon^{z_i^{+}z_j^{+}p_k}&=i\frac{\langle \Bar{i} \Bar{j}\rangle\left(\langle \Bar{i}\Bar{k}\rangle\langle k \Bar{j}\rangle+\langle \Bar{i} \Bar{j}\rangle p_k\right)}{2 p_i p_j},
\end{split}\quad
\begin{split}
\epsilon^{z_i^{-}p_j p_k}&=i\frac{\langle i\Bar{k}\rangle\left(\langle i j\rangle\langle k\Bar{j}\rangle-\langle k i\rangle p_j\right)+\langle i\Bar{j}\rangle \langle i j\rangle p_k}{2 p_i},\\
    \epsilon^{z_i^{+}p_j p_k}&=\Bar{i}\frac{\langle i\Bar{k}\rangle\left(\langle \Bar{i} j\rangle\langle k\Bar{j}\rangle-\langle k \Bar{i}\rangle p_j\right)+\langle i\Bar{j}\rangle \langle \Bar{i} j\rangle p_k}{2 p_i}.
    \end{split}
\end{equation}
We also used the two dimensional Schouten identity on many occasions which reads
\begin{align}
\delta^a_f\epsilon^{bc}+\delta^{b}_f\epsilon^{ac}+\delta^c_f\epsilon^{ba}=0.
\end{align}
Another very important manifestation of the Schouten identity is the following. For any two-component spinor $\lambda_{ia}$ we decompose it as a linear combination of $\lambda_{1a}$ and $\bar{\lambda}_{1a}$.
\begin{align}
    \lambda_{i}&=-\frac{\langle i\bar{j}\rangle}{2p_{j}}\lambda_{j}+\frac{\langle ij\rangle}{2p_{j}}\bar{\lambda}_{j},\\
\bar{\lambda}_{i}&=-\frac{\langle\bar{i}\bar{j}\rangle}{2p_{j}}\lambda_{j}+\frac{\langle\bar{i}j\rangle}{2p_{j}}\bar{\lambda}_{j}.
\end{align}
\section{ $\mathcal{N}=1$ and  $\mathcal{N}=2$ superconformal algebra}\label{sec:algbera}
\subsection*{ $\mathcal{N}=1$  superconformal algebra}
The $\mathcal{N}=1$ superconformal algebra in three dimensions consists of the following generators: The generator of translations, $P_\mu$, the generator of rotations, $M_{\mu\nu}$, the generator of dilatations and the generator of special conformal transformations which are respectively denoted as $D$ and $K_\mu$.
\begin{align}\label{SCFTalgebra1}
[&M_{\mu\nu},M_{\rho\lambda}]=i\left(\delta_{\mu\rho}M_{\nu\lambda}-\delta_{\nu\rho}M_{\mu\lambda}-\delta_{\mu\lambda}M_{\nu\rho}+\delta_{\nu\lambda}M_{\mu\rho}\right),\notag\\
    [M_{\mu\nu},P_\alpha]&=i\left(\delta_{\mu\alpha}P_\nu-\delta_{\nu\alpha}P_\mu\right),\qquad\quad[M_{\mu\nu},K_\alpha]=i\left(\delta_{\mu\alpha}K_\nu-\delta_{\nu\alpha}K_\mu\right),\notag\\
    [D,P_\mu]&=i P_\mu,\qquad \qquad\qquad\qquad~\quad[P_\mu,K_\nu]=2i\left(\delta_{\mu\nu}D-M_{\mu\nu}\right),\notag\\
    \{Q_a,Q_b\}&=i(\sigma^\mu)_{ab}P^\mu,\qquad\qquad\qquad~ \{S_a,S_b\}=i(\sigma^\mu)_{ab}K^\mu,\notag\\
    [D,Q_a]&=\frac{i}{2}Q_a,\qquad\qquad\qquad\qquad~~[D,K_\mu]=-i K_\mu,\notag\\
    [K_\mu,Q_a]&=i(\sigma_\mu)_a^b S_b,\qquad\qquad\qquad~~~[D,S_a]=-\frac{i}{2}S_a,\notag\\
    [M_{\mu\nu},Q_a]&=\frac{i}{2}\epsilon_{\mu\nu\rho}(\sigma^\rho)^b_a Q_b,\qquad\quad\quad~~[P_\mu,S_a]=i(\sigma_\mu)^a_b Q_b,
    \notag\\
     [M_{\mu\nu},S_a]&=\frac{i}{2}\epsilon_{\mu\nu\rho}(\sigma^\rho)^b_a S_b,\quad\qquad\quad~~\{Q_a,S_b\}=\epsilon_{ab}D-\frac{i}{2}\epsilon_{\mu\nu\rho}(\sigma^\rho)_{ab}M^{\mu\nu}.
\end{align}
The (anti) commutators not listed above are zero. The representation of these generators acting on primary superfields are as follows:\\
We have,
\begin{align}\label{posspacegen}
P_\mu&=-i\partial_\mu,\notag\\
M_{\mu\nu}&=-i\bigg(x_\mu\partial_\nu-x_\nu\partial_\mu+\frac{i}{2}\epsilon_{\mu\nu\rho}(\sigma^\rho)^{a}_b\theta^b\frac{\partial}{\partial\theta^a}\bigg)+\mathcal{M}_{\mu\nu},\notag\\
D&=-i\bigg(x^\nu\partial_\nu+\frac{1}{2}\theta^a\frac{\partial}{\partial\theta^a}+\Delta\bigg),\notag\\
K_\mu&=i\bigg(2x_\mu x^\nu \partial_\nu-x^2\partial_\mu+2\Delta x_\mu\bigg)-2x^\nu\mathcal{M}_{\mu\nu}+i(\sigma^\mu)^c_b(\sigma^\nu)^a_c x_\nu\theta^b\frac{\partial}{\partial \theta^a},\notag\\
Q_a&=\frac{\partial}{\partial\theta^a}+\frac{i}{2}\theta_b(\sigma^\mu)^b_{a}\partial_\mu,
\end{align}
where $\Delta$ is the scaling dimension of the primary and $\mathcal{M}_{\mu\nu}$ encodes the non trivial transformation of operators with spin. The expression for $S_a$ is obtained by computing $[K_\mu,Q_a]$ as in \eqref{SCFTalgebra1}. Given the position superspace generators in \eqref{posspacegen}, it is straightforward to obtain the momentum superspace generators by performing a Fourier transform to go from the $x^\mu$ to the $p_\mu$ variables. 

\begin{align}\label{momspacegen}
 P_\mu&=p_\mu,\notag\\
M_{\mu\nu}&=-i\bigg( p_\mu \frac{\partial}{\partial p_\nu}-p_\nu\frac{\partial}{\partial p_\mu}+\frac{i}{2}\epsilon_{\mu\nu\rho}(\sigma^\rho)^{a}_{b}\theta^b\frac{\partial}{\partial \theta^a}\bigg)+\mathcal{M}_{\mu\nu},\notag\\
    D&=i\bigg(p_\nu\frac{\partial}{\partial p_\nu}+(3-\Delta)-\frac{1}{2}\theta^a\frac{\partial}{\partial \theta^a}\bigg),\notag\\
    K_\mu&=-\bigg(p_\mu\frac{\partial^2}{\partial p_\nu\partial p^\nu}+2(\Delta-3)\frac{\partial}{\partial p_\mu}-2p_\nu\frac{\partial^2}{\partial p_\nu\partial p_\mu}\bigg)-2i\frac{\partial}{\partial p^\nu}\mathcal{M}_{\mu\nu}-(\sigma^\mu)^c_b(\sigma^\nu)^a_c\theta^b\frac{\partial}{\partial\theta^a}\frac{\partial}{\partial p^\nu},\notag\\
    Q_a&=\frac{\partial}{\partial\theta^a}-\frac{1}{2}\theta_b(\sigma^\mu)^b_{a}p_\mu,
    \end{align}
with $S_a$ obtained via \eqref{SCFTalgebra1}.\\
The generators in the super spinor helicity grassmann twistor variables \eqref{xipmdef} are given by,
\begin{align}\label{xispacegen}
    P_\mu&=\frac{1}{2}(\sigma^\mu)^a_b\lambda_a\,\overline{\lambda}^b,\notag\\
    M_{\mu\nu}&=\frac{1}{2}\epsilon_{\mu\nu\rho}(\sigma^\rho)^{a}_{b}\bigg(\,\overline{\lambda}^b\frac{\partial}{\partial\,\overline{\lambda}^a}+\lambda^b\frac{\partial}{\partial\lambda^a}\bigg),\notag\\
    D&=\frac{i}{2}\bigg(\,\overline{\lambda}^a\frac{\partial}{\partial\,\overline{\lambda}^a}+\lambda^a\frac{\partial}{\partial\lambda^a}+2\bigg),\notag\\
    K^\mu&=2(\sigma^\mu)^{ab}\frac{\partial^2}{\partial\lambda^a\,\overline{\lambda}^b},\notag\\
    Q_a&=\bigg(2\lambda_{a}\frac{\partial}{\partial \xi_+}+\frac{\,\overline{\lambda}_{a}}{4}\xi_+\bigg),\notag\\
    S_a&=-2i\bigg(2\frac{\partial}{\partial \xi_+}\frac{\partial}{\partial\,\overline{\lambda}^a}+\frac{\xi_+}{4}\frac{\partial}{\partial\lambda^a}\bigg).
\end{align}
\subsection*{$\mathcal{N}=2$ superconformal algebra}
Similarly we get the $\mathcal{N}=2$ algebra, all the generators will remain the same except the $Q_a$ and $S_a$ and we will have two more generators $\bar{Q}_a$ and $\bar{S}_a$. \\
$Q_a$ and $\bar{Q}_a$ were given in the Main text \eqref{xiomegaspacegen}, but we give them here agin with $S_a$ and $\bar{S}_a$.
\begin{equation}
    Q_a=2\lambda_a\frac{\partial}{\partial\omega_+}+\frac{\Bar{\lambda}_a}{4}\xi_+,\qquad
    \Bar{Q}_a=2\lambda_a\frac{\partial}{\partial\xi_+}+\frac{\Bar{\lambda}_a}{4}\omega_+,
\end{equation}
 \begin{equation}
      S_a=-2i\bigg(2\frac{\partial}{\partial\omega_+}\frac{\partial}{\partial\Bar{\lambda}^a}+\frac{\xi_+}{4}\frac{\partial}{\partial\lambda^a}\bigg), \quad\Bar{S}_a=-2i\bigg(2\frac{\partial}{\partial\xi_+}\frac{\partial}{\partial\Bar{\lambda}^a}+\frac{\omega_+}{4}\frac{\partial}{\partial \lambda^a}\bigg).
 \end{equation}
 For the sake of completeness we again write the R-symmetry generator \eqref{Roperator}
 \begin{align}
    R=\omega_+\frac{\partial}{\partial\omega_+}-\xi_+\frac{\partial}{\partial \xi_+}.
\end{align}

\section{$\mathbf{K_{SCFT}\implies K_{CFT}}$ for the components}\label{sec:KsusygivesK}
In this section we will prove that the invariance under special conformal transformations of the Super-correlators translates to the conformal invariance of the individual component correlators. Let us first consider an example viz $ \langle \mathbf{J}^{(ab)}\mathbf{J}^{(cd)}\mathbf{T}^{(ijkl)}\rangle$. Using the superfield expansion \eqref{Jsinthetamomspace}, we obtain,
\begin{equation}\label{JJTsuper}
    \langle \mathbf{J}^{(ab)}\mathbf{J}^{(cd)}\mathbf{T}^{(ijkl)}\rangle=\langle J^{ab}J^{cd}T^{ijkl}\rangle-\theta_{1e}\theta_{2f}\langle J^{abe}_{3/2}J^{cdf}_{3/2}T^{ijkl}\rangle +\dots
\end{equation}
The action of $K_\mu$ on a three point function is given by \footnote{The momentum space generators are provided in \eqref{momspacegen}.},
\begin{align}
     \notag K_\mu&=K_{1\mu}+K_{2\mu}\\
    \notag &=-\sum_{i=1}^2\bigg(p_{i\mu}\frac{\partial^2}{\partial p_{i\nu}\partial p^\nu_i}+2(\Delta_i-3)\frac{\partial}{\partial p_{i\mu}}-2p_{i\nu}\frac{\partial^2}{\partial p_{i\nu}\partial p_{i\mu}}\bigg)-\sum_{i=1}^22i\frac{\partial}{\partial p^\nu_i}\mathcal{M}_{i\mu\nu}\\
      &\quad  -\sum_{i=1}^2(\sigma^\mu)^c_b(\sigma^\nu)^a_c\theta^b_i\frac{\partial}{\partial\theta^a_i}\frac{\partial}{\partial p^\nu_i},
\end{align}
Acting with $K_\mu$ on \eqref{JJTsuper}, we get\footnote{Note that $K_\mu$ does not mix correlators appearing at different orders in the Grassmann spinor expansion. },
\begin{equation}\label{KmuonJJTsuper}
    K_{\mu} \langle \mathbf{J}^{(ab)}\mathbf{J}^{(cd)}\mathbf{T}^{(ijkl)}\rangle=K_{\mu}\langle J^{ab}J^{cd}T^{ijkl}\rangle-K_{\mu}\big(\theta_{1e}\theta_{2f}\langle J^{abe}_{3/2}J^{cdf}_{3/2}T^{ijkl}\rangle\big)+\cdots
\end{equation}
At the zeroth order of this Grassmann expansion, $K_\mu^\theta$(which we define below) doesn't play any role, but it will play a crucial role at higher orders of expansion in getting the correct special conformal transformation. We can write the action abstractly as,
\begin{align*}
    K_{\mu}\langle J^{ab}J^{cd}T\rangle&=K^{\langle JJT\rangle}_\mu\langle J^{ab}J^{cd}T\rangle+K_\mu^\theta \langle J^{ab}J^{cd}T\rangle,\\
    &=K^{\langle JJT\rangle}_\mu\langle J^{ab}J^{cd}T\rangle+0,\\
   \implies K_{\mu}\langle J^{ab}J^{cd}T\rangle &=K^{\langle JJT\rangle}_\mu\langle J^{ab}J^{cd}T\rangle,
\end{align*}
i.e, the usual conformal ward identity for the component correlator is reproduced.\\
where we write the first part of $K_\mu$ as $K^{\langle JJT\rangle}$, as its form is fixed by the zeroth component,
\begin{align}
K_\mu^{\langle JJT\rangle}&=-\sum_{i=1}^2\bigg(p_{i\mu}\frac{\partial^2}{\partial p_{i\nu}\partial p^\nu_i}+2(\Delta_i-3)\frac{\partial}{\partial p_{i\mu}}-2p_{i\nu}\frac{\partial^2}{\partial p_{i\nu}\partial p_{i\mu}}\bigg)-\sum_{i=1}^22i\frac{\partial}{\partial p^\nu_i}\mathcal{M}_{i\mu\nu}\label{KJJT},\\
K^{\theta}_\mu&=-\sum_{i=1}^2(\sigma^\mu)^c_b(\sigma^\nu)^a_c\theta^b_i\frac{\partial}{\partial\theta^a_i}\frac{\partial}{\partial p^\nu_i}.\label{Ktheta}
\end{align}
Now, we give the actual form of $K_{\mu}^{\langle JJT\rangle}$.
First, the action of $\mathcal{M}_{\mu\nu}$ on a spinor is given as 
\begin{equation}
    \mathcal{M}_{\mu\nu}F^a=-\frac{1}{2}\epsilon_{\mu\nu\rho}(\sigma^\rho)^a_b F^b.
\end{equation}
Now, let's evaluate the action of $\mathcal{M}_{\mu\nu}$ part on the zeroth order correlator.
\begin{align}
    -\sum_{i=1}^2 2i \frac{\partial}{\partial p_i^\nu} \mathcal{M}_{i\mu\nu}\langle J^{ab}J^{cd}T\rangle&=-2i\Big[\frac{\partial}{\partial p_1^\nu} \mathcal{M}_{1\mu\nu}\langle J^{ab}J^{cd}T\rangle+\frac{\partial}{\partial p_2^\nu} \mathcal{M}_{2\mu\nu}\langle J^{ab}J^{cd}T\rangle\Big]\notag\\
    &=-2i\bigg[-\frac{1}{2}\epsilon_{\mu\nu\rho}(\sigma^\rho)^a_k \frac{\partial}{\partial p_1^\nu}\langle J^{kb}J^{cd}T\rangle-\frac{1}{2}\epsilon_{\mu\nu\rho}(\sigma^\rho)^b_k \frac{\partial}{\partial p_1^\nu}\langle J^{ak}J^{cd}T\rangle\notag\\
    &~\,\qquad\quad-\frac{1}{2}\epsilon_{\mu\nu\rho}(\sigma^\rho)^c_k \frac{\partial}{\partial p_2^\nu}\langle J^{ab}J^{kd}T\rangle-\frac{1}{2}\epsilon_{\mu\nu\rho}(\sigma^\rho)^d_k \frac{\partial}{\partial p_2^\nu}\langle J^{ab}J^{ck}T\rangle\bigg]\label{CurlyMaction}
\end{align}
Hence, the representation of $\mathcal{M}_{\mu\nu}$ part is fixed, and it acts on the $a,b$ index of first insertion and on the $c,d$ index of second insertion. Thus, the whole expression can be given as 
\begin{align}\label{KJJTaction}
    K_\mu^{\langle JJT\rangle}\langle J^{ab}J^{cd}T\rangle=\left[-\sum_{i=1}^2\bigg(p_{i\mu}\frac{\partial^2}{\partial p_{i\nu}\partial p^\nu_i}+2(\Delta_i-3)\frac{\partial}{\partial p_{i\mu}}-2p_{i\nu}\frac{\partial^2}{\partial p_{i\nu}\partial p_{i\mu}}\bigg)-\sum_{i=1}^22i\frac{\partial}{\partial p^\nu_i}\mathcal{M}_{i\mu\nu}\right]\langle J^{ab}J^{cd}T\rangle,
\end{align}
where $\Delta_i=2$ i.e. the scaling dimension of spin-1 conserved currents, and the $\mathcal{M}_{\mu\nu}$ action is given by \eqref{CurlyMaction}. Hence, the action of $K^\mu_{\langle JJT \rangle}$ is fixed by the zeroth order, given in \eqref{KJJTaction}.\\
At the second order i.e. $O(\theta_1\theta_2)$,  we obtain, 
\begin{align}\label{secondorderKaction}
&K_{\mu}\big(\theta_{1e}\theta_{2f}\langle J^{abe}_{3/2}J^{cdf}_{3/2}T\rangle\big)=\theta_{1e}\theta_{2f} K_{\mu}^{\langle JJT\rangle}\langle J^{abe}_{3/2}J^{cdf}_{3/2}T\rangle+K^\mu_\theta \left(\theta_{1e}\theta_{2f} \langle J^{abe}_{3/2}J^{cdf}_{3/2}T\rangle\right),\notag\\
    &=\theta_{1e}\theta_{2f}\bigg[-\sum_{i=1}^2\bigg(p_{i\mu}\frac{\partial^2}{\partial p_{i\nu}\partial p^\nu_i}+2(\Delta_i-3)\frac{\partial}{\partial p_{i\mu}}-2p_{i\nu}\frac{\partial^2}{\partial p_{i\nu}\partial p_{i\mu}}\bigg)-\sum_{i=1}^22i\frac{\partial}{\partial p^\nu_i}\mathcal{M}_{i\mu\nu}\bigg]\langle J^{abe}_{3/2}J^{cdf}_{3/2}T\rangle\notag\\
      &~~\,\quad\quad  -\bigg[\sum_{i=1}^2(\sigma^\mu)^c_b(\sigma^\nu)^a_c\theta^b_i\frac{\partial}{\partial\theta^a_i}\frac{\partial}{\partial p^\nu_i}\bigg]\left(\theta_{1e}\theta_{2f}\langle J^{abe}_{3/2}J^{cdf}_{3/2}T\rangle\right),
\end{align}
where $\Delta_i=2$ are scaling dimensions of the first two insertions in $\langle JJT\rangle$. \\
Consider now the third of line of \eqref{secondorderKaction}. We have,
\begin{align}
K_\mu^{\theta}&\left(\theta_{1e}\theta_{2f}\langle J^{abe}_{3/2}J^{cdf}_{3/2}T\rangle\right)=-\bigg[\sum_{i=1}^2(\sigma^\mu)^c_b(\sigma^\nu)^a_c\theta^b_i\frac{\partial}{\partial\theta^a_i}\frac{\partial}{\partial p^\nu_i}\bigg](\theta_{1e}\theta_{2f})\langle J^{abe}_{3/2}J^{cdf}_{3/2}T\rangle,\notag\\
     &=-\Big(\delta^{\mu\nu}\delta^l_k+i \epsilon^{\mu\nu\rho}(\sigma^\rho)^l_k\Big)\left(\theta_1^k\theta_{2f}\epsilon_{le}\frac{\partial}{\partial p_1^\nu}\langle J^{abe}_{3/2}J^{cdf}_{3/2}T\rangle+\theta_{1e}\theta_2^k \epsilon_{lf}\frac{\partial}{\partial p_2^\nu}\langle J^{abe}_{3/2}J^{cdf}_{3/2}T\rangle\right),\notag\\
&=-\theta_{1e}\theta_{2f}\Big[\frac{\partial}{\partial p_1^\mu}\langle J^{abe}_{3/2}J^{cdf}_{3/2}T\rangle+\frac{\partial}{\partial p_2^\mu}\langle J^{abe}_{3/2}J^{cdf}_{3/2}T\rangle\Big]\notag\\
&\quad+\theta_{1e}\theta_{2f}\left[i\epsilon^{\mu\nu\rho}(\sigma_\rho)^e_k \frac{\partial}{\partial p_1^\nu}\langle J^{abk}_{3/2}J^{cdf}_{3/2}T\rangle+i\epsilon^{\mu\nu\rho}(\sigma_\rho)^f_k \frac{\partial}{\partial p_2^\nu}\langle J^{abe}_{3/2}J^{cdk}_{3/2}T\rangle\right],\notag\\
&=-2\theta_{1e}\theta_{2f}\Big[\frac{1}{2}\frac{\partial}{\partial p_1^\mu}\langle J^{abe}_{3/2}J^{cdf}_{3/2}T\rangle+\frac{1}{2}\frac{\partial}{\partial p_2^\mu}\langle J^{abe}_{3/2}J^{cdf}_{3/2}T\rangle\Big]\notag\\
&\quad-2\theta_{1e}\theta_{2f}\left[-\frac{1}{2}\epsilon^{\mu\nu\rho}(\sigma_\rho)^e_k \frac{\partial}{\partial p_1^\nu}\langle J^{abk}_{3/2}J^{cdf}_{3/2}T\rangle-\frac{1}{2}\epsilon^{\mu\nu\rho}(\sigma_\rho)^f_k\frac{\partial}{\partial p_2^\nu} \langle J^{abe}_{3/2}J^{cdk}_{3/2}T\rangle\right].
\end{align}
Hence, 
\begin{align} \label{Kthetaaction}
K_\mu^{\theta}\left(\theta_{1e}\theta_{2f}\langle J^{abe}_{3/2}J^{cdf}_{3/2}T\rangle\right)&=-2\theta_{1e}\theta_{2f}\Big[\frac{1}{2}\frac{\partial}{\partial p_1^\mu}\langle J^{abe}_{3/2}J^{cdf}_{3/2}T\rangle+\frac{1}{2}\frac{\partial}{\partial p_2^\mu}\langle J^{abe}_{3/2}J^{cdf}_{3/2}T\rangle\Big]\notag\\
&\quad-2\theta_{1e}\theta_{2f}\left[-\frac{1}{2}\epsilon^{\mu\nu\rho}(\sigma_\rho)^e_k \frac{\partial}{\partial p_1^\nu}\langle J^{abk}_{3/2}J^{cdf}_{3/2}T\rangle-\frac{1}{2}\epsilon^{\mu\nu\rho}(\sigma_\rho)^f_k\frac{\partial}{\partial p_2^\nu} \langle J^{abe}_{3/2}J^{cdk}_{3/2}T\rangle\right].
\end{align}
Using \eqref{Kthetaaction} in \eqref{secondorderKaction}, it becomes, 
\begin{align}
    &K_{\mu}\big(\theta_{1e}\theta_{2f}\langle J^{abe}_{3/2}J^{cdf}_{3/2}T\rangle\big)=\theta_{1e}\theta_{2f} K_{\mu}^{\langle JJT\rangle}\langle J^{abe}_{3/2}J^{cdf}_{3/2}T\rangle+K_\mu^\theta \left(\theta_{1e}\theta_{2f} \langle J^{abe}_{3/2}J^{cdf}_{3/2}T\rangle\right),\notag\\
    &=\theta_{1e}\theta_{2f}\bigg[-\sum_{i=1}^2\bigg(p_{i\mu}\frac{\partial^2}{\partial p_{i\nu}\partial p^\nu_i}+2\left(\Delta_i+\frac{1}{2}-3\right)\frac{\partial}{\partial p_{i\mu}}-2p_{i\nu}\frac{\partial^2}{\partial p_{i\nu}\partial p_{i\mu}}\bigg)-\sum_{i=1}^22i\frac{\partial}{\partial p^\nu_i}\mathcal{M}_{i\mu\nu}\bigg]\langle J^{abe}_{3/2}J^{cdf}_{3/2}T\rangle,\notag\\
    &=\theta_{1e}\theta_{2f}\Big(K_\mu^{\langle J_{3/2}J_{3/2}T \rangle}\langle J^{abe}_{3/2}J^{cdf}_{3/2}T\rangle\Big).
\end{align}
Hence, after the addition of $K_\mu^\theta$ action to $K^{\langle JJT\rangle}_\mu$, we get the correct action on $\langle J_{3/2}J_{3/2}T\rangle$. This can be seen as the dimensions have shifted, $\Delta_i \to \Delta_i+\frac{1}{2}$, which are the correct scaling dimension of spin-3/2 conserved current. Also, the correct action of the index $e$ and $f$ for both first and second insertion is added to the $\mathcal{M}_{\mu\nu}$ part. Thus making it the special conformal transformation for the component correlator. This method can be applied to any component. Hence, we have shown that the invariance under special conformal transformations of the Super-correlators translates to the conformal invariance of the individual component correlators
\section{A worked out example of a $\mathcal{N}=1$ three point function}\label{sec:example}
In this appendix, we outline the steps for solving for $\mathcal{N}=1$ three-point correlators in the Grassmann twistor variables, demonstrating the significant reduction of work in getting the constraints between different component correlator. Consider the Super-correlator $\langle\tilde{\mathbf{J}}_{3/2}\tilde{\mathbf{J}}_{1/2}\tilde{\mathbf{J}}_{1/2}\rangle$. We have, (see equation \eqref{threepoint3halfInt})
\begin{align}
    \langle\tilde{\mathbf{J}}^-_{3/2} \tilde{\mathbf{J}}^-_{1/2} \tilde{\mathbf{J}}^-_{1/2}\rangle&=\frac{ E}{32\langle\overline{1}\,\overline{2}\rangle\sqrt{p_3}}\,\langle J^-_{3/2}O^-_{1/2}J^-\rangle\,\Xi_3,\label{mmm}\\
\langle\tilde{\mathbf{J}}^-_{3/2} \tilde{\mathbf{J}}^-_{1/2} \tilde{\mathbf{J}}^+_{1/2}\rangle&=-\frac{\xi_{3-} }{256\sqrt{p_3}}\,\langle J^-_{3/2}O^-_{1/2}J^+\rangle\,\Gamma_3\label{mmp}.
\end{align}
In equation \eqref{mmm}, we can extract the component correlators on the LHS using the superfield expansion \eqref{JsinXipm} and compare with the corresponding expression in the RHS. Let's first consider \eqref{mmm}. By matching the LHS and RHS at the zeroth order, $\xi_{2+}\xi_{3+}$, and $\xi_{1+}\xi_{3+}$, we get the following relations
\begin{align}
    \langle T^-J^-J^-\rangle&=\frac{2 E\sqrt{p_1p_2}}{\langle\Bar{1}\Bar{2}\rangle}\langle J^-_{3/2}O^-_{1/2}J^-\rangle,\\
     \langle T^-O_{1/2}^-O_{1/2}^-\rangle&=\frac{ \langle\Bar{2}\Bar{3}\rangle\sqrt{p_1}}{\langle\Bar{1}\Bar{2}\rangle\sqrt{p_3}}\langle J^-_{3/2}O^-_{1/2}J^-\rangle,\\
      \langle J_{3/2}^-J^-O_{1/2}^-\rangle &=\frac{\langle\Bar{1}\Bar{3}\rangle\sqrt{p_2}}{\langle\Bar{1}\Bar{2}\rangle\sqrt{p_3}}\langle J^-_{3/2}O^-_{1/2}J^-\rangle.
\end{align}
This implies that,
\begin{align}\label{mmmEQ}
     \langle T^-J^-J^-\rangle=\frac{2 E\sqrt{p_2p_3}}{\langle\Bar{2}\Bar{3}\rangle}\langle T^-O^-_{1/2}O_{1/2}^-\rangle,\quad \langle T^-J^-J^-\rangle=\frac{2 E\sqrt{p_1p_3}}{\langle\Bar{1}\Bar{3}\rangle}\langle J_{3/2}^-J^-O_{1/2}^-\rangle.
\end{align}
Similarly, in the for $(- - +)$ helicity \eqref{mmp} we get the following relations between the component correlators:
\begin{equation}\label{mmpEQ}
     \langle T^-J^-J^+\rangle=2 \sqrt{p_2p_3}\frac{\langle12\rangle}{\langle 31\rangle}\langle T^-O^-_{1/2}O_{1/2}^+\rangle,\quad \langle T^-J^-J^+\rangle=2\sqrt{p_1p_3} \frac{\langle12\rangle}{\langle23\rangle}\langle J_{3/2}^-J^-O_{1/2}^+\rangle.
\end{equation}
Using the decomposition of a three dimensional CFT correlator as in \eqref{s1s2s3comp}, we can write explicit forms for these correlators. Substituting these forms in equations \eqref{mmmEQ} and \eqref{mmpEQ}, we can constrain their individual coefficients and reduce the independent coefficients from six to one. Then, correlators can be written as
\begin{align}
     \langle TJJ\rangle&=n_b\big(\langle TJJ\rangle_B+\langle TJJ\rangle_F\big),\\
    \langle TO_{1/2}O_{1/2}\rangle&=n_b\big(\langle TO_{1/2}O_{1/2}\rangle_B+\langle TO_{1/2}O_{1/2}\rangle_F\big),\\
    \langle J_{1/2}JO_{1/2}\rangle&=n_b\big(\langle J_{1/2}JO_{1/2}\rangle_B+\langle J_{1/2}JO_{1/2}\rangle_F\big).
\end{align}
    Thus, we obtain exactly one parity even solution which is exactly in line with the results of \cite{Buchbinder:2023ndg}.
\bibliographystyle{JHEP}
\bibliography{biblio}

\providecommand{\href}[2]{#2}\begingroup\raggedright\begin{thebibliography}{10}

\bibitem{Osborn:1998qu}
H.~Osborn, \emph{{N=1 superconformal symmetry in four-dimensional quantum field theory}}, \href{https://doi.org/10.1006/aphy.1998.5893}{\emph{Annals Phys.} {\bfseries 272} (1999) 243} [\href{https://arxiv.org/abs/hep-th/9808041}{{\ttfamily hep-th/9808041}}].

\bibitem{Park:1999pd}
J.-H.~Park, \emph{{Superconformal symmetry and correlation functions}}, \href{https://doi.org/10.1016/S0550-3213(99)00432-0}{\emph{Nucl. Phys. B} {\bfseries 559} (1999) 455} [\href{https://arxiv.org/abs/hep-th/9903230}{{\ttfamily hep-th/9903230}}].

\bibitem{Maldacena:2011nz}
J.M.~Maldacena and G.L.~Pimentel, \emph{{On graviton non-Gaussianities during inflation}}, \href{https://doi.org/10.1007/JHEP09(2011)045}{\emph{JHEP} {\bfseries 09} (2011) 045} [\href{https://arxiv.org/abs/1104.2846}{{\ttfamily 1104.2846}}].

\bibitem{Raju:2012zr}
S.~Raju, \emph{{New Recursion Relations and a Flat Space Limit for AdS/CFT Correlators}}, \href{https://doi.org/10.1103/PhysRevD.85.126009}{\emph{Phys. Rev. D} {\bfseries 85} (2012) 126009} [\href{https://arxiv.org/abs/1201.6449}{{\ttfamily 1201.6449}}].

\bibitem{Farrow:2018yni}
J.A.~Farrow, A.E.~Lipstein and P.~McFadden, \emph{{Double copy structure of CFT correlators}}, \href{https://doi.org/10.1007/JHEP02(2019)130}{\emph{JHEP} {\bfseries 02} (2019) 130} [\href{https://arxiv.org/abs/1812.11129}{{\ttfamily 1812.11129}}].

\bibitem{Lipstein:2019mpu}
A.E.~Lipstein and P.~McFadden, \emph{{Double copy structure and the flat space limit of conformal correlators in even dimensions}}, \href{https://doi.org/10.1103/PhysRevD.101.125006}{\emph{Phys. Rev. D} {\bfseries 101} (2020) 125006} [\href{https://arxiv.org/abs/1912.10046}{{\ttfamily 1912.10046}}].

\bibitem{Jain:2021qcl}
S.~Jain, R.R.~John, A.~Mehta, A.A.~Nizami and A.~Suresh, \emph{{Double copy structure of parity-violating CFT correlators}}, \href{https://doi.org/10.1007/JHEP07(2021)033}{\emph{JHEP} {\bfseries 07} (2021) 033} [\href{https://arxiv.org/abs/2104.12803}{{\ttfamily 2104.12803}}].

\bibitem{McFadden:2011kk}
P.~McFadden and K.~Skenderis, \emph{{Cosmological 3-point correlators from holography}}, \href{https://doi.org/10.1088/1475-7516/2011/06/030}{\emph{JCAP} {\bfseries 06} (2011) 030} [\href{https://arxiv.org/abs/1104.3894}{{\ttfamily 1104.3894}}].

\bibitem{Ghosh:2014kba}
A.~Ghosh, N.~Kundu, S.~Raju and S.P.~Trivedi, \emph{{Conformal Invariance and the Four Point Scalar Correlator in Slow-Roll Inflation}}, \href{https://doi.org/10.1007/JHEP07(2014)011}{\emph{JHEP} {\bfseries 07} (2014) 011} [\href{https://arxiv.org/abs/1401.1426}{{\ttfamily 1401.1426}}].

\bibitem{Arkani-Hamed:2015bza}
N.~Arkani-Hamed and J.~Maldacena, \emph{{Cosmological Collider Physics}},  \href{https://arxiv.org/abs/1503.08043}{{\ttfamily 1503.08043}}.

\bibitem{Arkani-Hamed:2018kmz}
N.~Arkani-Hamed, D.~Baumann, H.~Lee and G.L.~Pimentel, \emph{{The Cosmological Bootstrap: Inflationary Correlators from Symmetries and Singularities}}, \href{https://doi.org/10.1007/JHEP04(2020)105}{\emph{JHEP} {\bfseries 04} (2020) 105} [\href{https://arxiv.org/abs/1811.00024}{{\ttfamily 1811.00024}}].

\bibitem{Baumann:2022jpr}
D.~Baumann, D.~Green, A.~Joyce, E.~Pajer, G.L.~Pimentel, C.~Sleight et~al., \emph{{Snowmass White Paper: The Cosmological Bootstrap}},  in \emph{{Snowmass 2021}}, 3, 2022 [\href{https://arxiv.org/abs/2203.08121}{{\ttfamily 2203.08121}}].

\bibitem{Raju:2010by}
S.~Raju, \emph{{BCFW for Witten Diagrams}}, \href{https://doi.org/10.1103/PhysRevLett.106.091601}{\emph{Phys. Rev. Lett.} {\bfseries 106} (2011) 091601} [\href{https://arxiv.org/abs/1011.0780}{{\ttfamily 1011.0780}}].

\bibitem{Raju:2012zs}
S.~Raju, \emph{{Four Point Functions of the Stress Tensor and Conserved Currents in AdS$_4$/CFT$_3$}}, \href{https://doi.org/10.1103/PhysRevD.85.126008}{\emph{Phys. Rev. D} {\bfseries 85} (2012) 126008} [\href{https://arxiv.org/abs/1201.6452}{{\ttfamily 1201.6452}}].

\bibitem{Albayrak:2018tam}
S.~Albayrak and S.~Kharel, \emph{{Towards the higher point holographic momentum space amplitudes}}, \href{https://doi.org/10.1007/JHEP02(2019)040}{\emph{JHEP} {\bfseries 02} (2019) 040} [\href{https://arxiv.org/abs/1810.12459}{{\ttfamily 1810.12459}}].

\bibitem{Albayrak:2019yve}
S.~Albayrak and S.~Kharel, \emph{{Towards the higher point holographic momentum space amplitudes. Part II. Gravitons}}, \href{https://doi.org/10.1007/JHEP12(2019)135}{\emph{JHEP} {\bfseries 12} (2019) 135} [\href{https://arxiv.org/abs/1908.01835}{{\ttfamily 1908.01835}}].

\bibitem{Bissi:2022wuh}
A.~Bissi, G.~Fardelli, A.~Manenti and X.~Zhou, \emph{{Spinning correlators in $ \mathcal{N} $ = 2 SCFTs: Superspace and AdS amplitudes}}, \href{https://doi.org/10.1007/JHEP01(2023)021}{\emph{JHEP} {\bfseries 01} (2023) 021} [\href{https://arxiv.org/abs/2209.01204}{{\ttfamily 2209.01204}}].

\bibitem{Coriano:2013jba}
C.~Coriano, L.~Delle~Rose, E.~Mottola and M.~Serino, \emph{{Solving the Conformal Constraints for Scalar Operators in Momentum Space and the Evaluation of Feynman's Master Integrals}}, \href{https://doi.org/10.1007/JHEP07(2013)011}{\emph{JHEP} {\bfseries 07} (2013) 011} [\href{https://arxiv.org/abs/1304.6944}{{\ttfamily 1304.6944}}].

\bibitem{Bzowski:2013sza}
A.~Bzowski, P.~McFadden and K.~Skenderis, \emph{{Implications of conformal invariance in momentum space}}, \href{https://doi.org/10.1007/JHEP03(2014)111}{\emph{JHEP} {\bfseries 03} (2014) 111} [\href{https://arxiv.org/abs/1304.7760}{{\ttfamily 1304.7760}}].

\bibitem{Bzowski:2015pba}
A.~Bzowski, P.~McFadden and K.~Skenderis, \emph{{Scalar 3-point functions in CFT: renormalisation, beta functions and anomalies}}, \href{https://doi.org/10.1007/JHEP03(2016)066}{\emph{JHEP} {\bfseries 03} (2016) 066} [\href{https://arxiv.org/abs/1510.08442}{{\ttfamily 1510.08442}}].

\bibitem{Bzowski:2017poo}
A.~Bzowski, P.~McFadden and K.~Skenderis, \emph{{Renormalised 3-point functions of stress tensors and conserved currents in CFT}}, \href{https://doi.org/10.1007/JHEP11(2018)153}{\emph{JHEP} {\bfseries 11} (2018) 153} [\href{https://arxiv.org/abs/1711.09105}{{\ttfamily 1711.09105}}].

\bibitem{Bzowski:2018fql}
A.~Bzowski, P.~McFadden and K.~Skenderis, \emph{{Renormalised CFT 3-point functions of scalars, currents and stress tensors}}, \href{https://doi.org/10.1007/JHEP11(2018)159}{\emph{JHEP} {\bfseries 11} (2018) 159} [\href{https://arxiv.org/abs/1805.12100}{{\ttfamily 1805.12100}}].

\bibitem{Bautista:2019qxj}
T.~Bautista and H.~Godazgar, \emph{{Lorentzian CFT 3-point functions in momentum space}}, \href{https://doi.org/10.1007/JHEP01(2020)142}{\emph{JHEP} {\bfseries 01} (2020) 142} [\href{https://arxiv.org/abs/1908.04733}{{\ttfamily 1908.04733}}].

\bibitem{Jain:2020rmw}
S.~Jain, R.R.~John and V.~Malvimat, \emph{{Momentum space spinning correlators and higher spin equations in three dimensions}}, \href{https://doi.org/10.1007/JHEP11(2020)049}{\emph{JHEP} {\bfseries 11} (2020) 049} [\href{https://arxiv.org/abs/2005.07212}{{\ttfamily 2005.07212}}].

\bibitem{Jain:2020puw}
S.~Jain, R.R.~John and V.~Malvimat, \emph{{Constraining momentum space correlators using slightly broken higher spin symmetry}}, \href{https://doi.org/10.1007/JHEP04(2021)231}{\emph{JHEP} {\bfseries 04} (2021) 231} [\href{https://arxiv.org/abs/2008.08610}{{\ttfamily 2008.08610}}].

\bibitem{Jain:2021wyn}
S.~Jain, R.R.~John, A.~Mehta, A.A.~Nizami and A.~Suresh, \emph{{Momentum space parity-odd CFT 3-point functions}}, \href{https://doi.org/10.1007/JHEP08(2021)089}{\emph{JHEP} {\bfseries 08} (2021) 089} [\href{https://arxiv.org/abs/2101.11635}{{\ttfamily 2101.11635}}].

\bibitem{Jain:2021vrv}
S.~Jain, R.R.~John, A.~Mehta, A.A.~Nizami and A.~Suresh, \emph{{Higher spin 3-point functions in 3d CFT using spinor-helicity variables}}, \href{https://doi.org/10.1007/JHEP09(2021)041}{\emph{JHEP} {\bfseries 09} (2021) 041} [\href{https://arxiv.org/abs/2106.00016}{{\ttfamily 2106.00016}}].

\bibitem{Jain:2021gwa}
S.~Jain and R.R.~John, \emph{{Relation between parity-even and parity-odd CFT correlation functions in three dimensions}}, \href{https://doi.org/10.1007/JHEP12(2021)067}{\emph{JHEP} {\bfseries 12} (2021) 067} [\href{https://arxiv.org/abs/2107.00695}{{\ttfamily 2107.00695}}].

\bibitem{Jain:2021whr}
S.~Jain, R.R.~John, A.~Mehta and D.K.~S, \emph{{Constraining momentum space CFT correlators with consistent position space OPE limit and the collider bound}}, \href{https://doi.org/10.1007/JHEP02(2022)084}{\emph{JHEP} {\bfseries 02} (2022) 084} [\href{https://arxiv.org/abs/2111.08024}{{\ttfamily 2111.08024}}].

\bibitem{Isono:2019ihz}
H.~Isono, T.~Noumi and T.~Takeuchi, \emph{{Momentum space conformal three-point functions of conserved currents and a general spinning operator}}, \href{https://doi.org/10.1007/JHEP05(2019)057}{\emph{JHEP} {\bfseries 05} (2019) 057} [\href{https://arxiv.org/abs/1903.01110}{{\ttfamily 1903.01110}}].

\bibitem{Gillioz:2019lgs}
M.~Gillioz, \emph{{Conformal 3-point functions and the Lorentzian OPE in momentum space}}, \href{https://doi.org/10.1007/s00220-020-03836-8}{\emph{Commun. Math. Phys.} {\bfseries 379} (2020) 227} [\href{https://arxiv.org/abs/1909.00878}{{\ttfamily 1909.00878}}].

\bibitem{Baumann:2019oyu}
D.~Baumann, C.~Duaso~Pueyo, A.~Joyce, H.~Lee and G.L.~Pimentel, \emph{{The cosmological bootstrap: weight-shifting operators and scalar seeds}}, \href{https://doi.org/10.1007/JHEP12(2020)204}{\emph{JHEP} {\bfseries 12} (2020) 204} [\href{https://arxiv.org/abs/1910.14051}{{\ttfamily 1910.14051}}].

\bibitem{Inbasekar:2015tsa}
K.~Inbasekar, S.~Jain, S.~Mazumdar, S.~Minwalla, V.~Umesh and S.~Yokoyama, \emph{{Unitarity, crossing symmetry and duality in the scattering of $ \mathcal{N}=1 $ susy matter Chern-Simons theories}}, \href{https://doi.org/10.1007/JHEP10(2015)176}{\emph{JHEP} {\bfseries 10} (2015) 176} [\href{https://arxiv.org/abs/1505.06571}{{\ttfamily 1505.06571}}].

\bibitem{Gur-Ari:2015pca}
G.~Gur-Ari and R.~Yacoby, \emph{{Three Dimensional Bosonization From Supersymmetry}}, \href{https://doi.org/10.1007/JHEP11(2015)013}{\emph{JHEP} {\bfseries 11} (2015) 013} [\href{https://arxiv.org/abs/1507.04378}{{\ttfamily 1507.04378}}].

\bibitem{Aharony:2019mbc}
O.~Aharony and A.~Sharon, \emph{{Large N renormalization group flows in 3d $ \mathcal{N} $ = 1 Chern-Simons-Matter theories}}, \href{https://doi.org/10.1007/JHEP07(2019)160}{\emph{JHEP} {\bfseries 07} (2019) 160} [\href{https://arxiv.org/abs/1905.07146}{{\ttfamily 1905.07146}}].

\bibitem{Aharony:2008ug}
O.~Aharony, O.~Bergman, D.L.~Jafferis and J.~Maldacena, \emph{{N=6 superconformal Chern-Simons-matter theories, M2-branes and their gravity duals}}, \href{https://doi.org/10.1088/1126-6708/2008/10/091}{\emph{JHEP} {\bfseries 10} (2008) 091} [\href{https://arxiv.org/abs/0806.1218}{{\ttfamily 0806.1218}}].

\bibitem{Nizami:2013tpa}
A.A.~Nizami, T.~Sharma and V.~Umesh, \emph{{Superspace formulation and correlation functions of 3d superconformal field theories}}, \href{https://doi.org/10.1007/JHEP07(2014)022}{\emph{JHEP} {\bfseries 07} (2014) 022} [\href{https://arxiv.org/abs/1308.4778}{{\ttfamily 1308.4778}}].

\bibitem{Buchbinder:2021gwu}
E.I.~Buchbinder and B.J.~Stone, \emph{{Mixed three-point functions of conserved currents in three-dimensional superconformal field theory}}, \href{https://doi.org/10.1103/PhysRevD.103.086023}{\emph{Phys. Rev. D} {\bfseries 103} (2021) 086023} [\href{https://arxiv.org/abs/2102.04827}{{\ttfamily 2102.04827}}].

\bibitem{Buchbinder:2021izb}
E.I.~Buchbinder, J.~Hutomo and S.M.~Kuzenko, \emph{{Correlation functions of spinor current multiplets in $ \mathcal{N} $ = 1 superconformal theory}}, \href{https://doi.org/10.1007/JHEP07(2021)165}{\emph{JHEP} {\bfseries 07} (2021) 165} [\href{https://arxiv.org/abs/2103.09472}{{\ttfamily 2103.09472}}].

\bibitem{Buchbinder:2021kjk}
E.I.~Buchbinder, J.~Hutomo and S.M.~Kuzenko, \emph{{Three-point functions of higher-spin spinor current multiplets in $ \mathcal{N} $ = 1 superconformal theory}}, \href{https://doi.org/10.1007/JHEP10(2021)058}{\emph{JHEP} {\bfseries 10} (2021) 058} [\href{https://arxiv.org/abs/2106.14498}{{\ttfamily 2106.14498}}].

\bibitem{Buchbinder:2021qlb}
E.I.~Buchbinder and B.J.~Stone, \emph{{Three-point functions of a superspin-2 current multiplet in 3D, N=1 superconformal theory}}, \href{https://doi.org/10.1103/PhysRevD.104.106004}{\emph{Phys. Rev. D} {\bfseries 104} (2021) 106004} [\href{https://arxiv.org/abs/2108.01865}{{\ttfamily 2108.01865}}].

\bibitem{Jain:2022izp}
A.~Jain and A.A.~Nizami, \emph{{Superconformal invariants and spinning correlators in 3d${{{\mathcal {N}}}}=2$ SCFTs}}, \href{https://doi.org/10.1140/epjc/s10052-022-11016-2}{\emph{Eur. Phys. J. C} {\bfseries 82} (2022) 1065} [\href{https://arxiv.org/abs/2205.11157}{{\ttfamily 2205.11157}}].

\bibitem{Buchbinder:2023fqv}
E.I.~Buchbinder and B.J.~Stone, \emph{{Three-point functions of conserved supercurrents in 3D N=1 SCFT: General formalism for arbitrary superspins}}, \href{https://doi.org/10.1103/PhysRevD.107.106001}{\emph{Phys. Rev. D} {\bfseries 107} (2023) 106001} [\href{https://arxiv.org/abs/2302.00593}{{\ttfamily 2302.00593}}].

\bibitem{Buchbinder:2023ndg}
E.I.~Buchbinder and B.J.~Stone, \emph{{Grassmann-odd three-point functions of conserved supercurrents in 3D N=1 SCFT}}, \href{https://doi.org/10.1103/PhysRevD.108.046001}{\emph{Phys. Rev. D} {\bfseries 108} (2023) 046001} [\href{https://arxiv.org/abs/2305.02233}{{\ttfamily 2305.02233}}].

\bibitem{Park:1999cw}
J.-H.~Park, \emph{{Superconformal symmetry in three-dimensions}}, \href{https://doi.org/10.1063/1.1290056}{\emph{J. Math. Phys.} {\bfseries 41} (2000) 7129} [\href{https://arxiv.org/abs/hep-th/9910199}{{\ttfamily hep-th/9910199}}].

\bibitem{Giombi:2011rz}
S.~Giombi, S.~Prakash and X.~Yin, \emph{{A Note on CFT Correlators in Three Dimensions}}, \href{https://doi.org/10.1007/JHEP07(2013)105}{\emph{JHEP} {\bfseries 07} (2013) 105} [\href{https://arxiv.org/abs/1104.4317}{{\ttfamily 1104.4317}}].

\bibitem{Witten:2003nn}
E.~Witten, \emph{{Perturbative gauge theory as a string theory in twistor space}}, \href{https://doi.org/10.1007/s00220-004-1187-3}{\emph{Commun. Math. Phys.} {\bfseries 252} (2004) 189} [\href{https://arxiv.org/abs/hep-th/0312171}{{\ttfamily hep-th/0312171}}].

\bibitem{Elvang:2013cua}
H.~Elvang and Y.-t.~Huang, \emph{{Scattering Amplitudes}},  \href{https://arxiv.org/abs/1308.1697}{{\ttfamily 1308.1697}}.

\bibitem{Elvang:2011fx}
H.~Elvang, Y.-t.~Huang and C.~Peng, \emph{{On-shell superamplitudes in N\ensuremath{<}4 SYM}}, \href{https://doi.org/10.1007/JHEP09(2011)031}{\emph{JHEP} {\bfseries 09} (2011) 031} [\href{https://arxiv.org/abs/1102.4843}{{\ttfamily 1102.4843}}].

\bibitem{Adamo:2022dcm}
T.~Adamo, J.J.M.~Carrasco, M.~Carrillo-Gonz\'alez, M.~Chiodaroli, H.~Elvang, H.~Johansson et~al., \emph{{Snowmass White Paper: the Double Copy and its Applications}},  in \emph{{Snowmass 2021}}, 4, 2022 [\href{https://arxiv.org/abs/2204.06547}{{\ttfamily 2204.06547}}].

\bibitem{jain2024scft}
\emph{To appear},  ~.

\bibitem{Maldacena:2012sf}
J.~Maldacena and A.~Zhiboedov, \emph{{Constraining conformal field theories with a slightly broken higher spin symmetry}}, \href{https://doi.org/10.1088/0264-9381/30/10/104003}{\emph{Class. Quant. Grav.} {\bfseries 30} (2013) 104003} [\href{https://arxiv.org/abs/1204.3882}{{\ttfamily 1204.3882}}].

\bibitem{Maldacena:2011jn}
J.~Maldacena and A.~Zhiboedov, \emph{{Constraining Conformal Field Theories with A Higher Spin Symmetry}}, \href{https://doi.org/10.1088/1751-8113/46/21/214011}{\emph{J. Phys. A} {\bfseries 46} (2013) 214011} [\href{https://arxiv.org/abs/1112.1016}{{\ttfamily 1112.1016}}].

\bibitem{Giombi:2016zwa}
S.~Giombi, V.~Gurucharan, V.~Kirilin, S.~Prakash and E.~Skvortsov, \emph{{On the Higher-Spin Spectrum in Large N Chern-Simons Vector Models}}, \href{https://doi.org/10.1007/JHEP01(2017)058}{\emph{JHEP} {\bfseries 01} (2017) 058} [\href{https://arxiv.org/abs/1610.08472}{{\ttfamily 1610.08472}}].

\bibitem{Turiaci:2018nua}
G.J.~Turiaci and A.~Zhiboedov, \emph{{Veneziano Amplitude of Vasiliev Theory}}, \href{https://doi.org/10.1007/JHEP10(2018)034}{\emph{JHEP} {\bfseries 10} (2018) 034} [\href{https://arxiv.org/abs/1802.04390}{{\ttfamily 1802.04390}}].

\bibitem{Aharony:2018npf}
O.~Aharony, L.F.~Alday, A.~Bissi and R.~Yacoby, \emph{{The Analytic Bootstrap for Large $N$ Chern-Simons Vector Models}}, \href{https://doi.org/10.1007/JHEP08(2018)166}{\emph{JHEP} {\bfseries 08} (2018) 166} [\href{https://arxiv.org/abs/1805.04377}{{\ttfamily 1805.04377}}].

\bibitem{Skvortsov:2018uru}
E.~Skvortsov, \emph{{Light-Front Bootstrap for Chern-Simons Matter Theories}}, \href{https://doi.org/10.1007/JHEP06(2019)058}{\emph{JHEP} {\bfseries 06} (2019) 058} [\href{https://arxiv.org/abs/1811.12333}{{\ttfamily 1811.12333}}].

\bibitem{Li:2019twz}
Z.~Li, \emph{{Bootstrapping conformal four-point correlators with slightly broken higher spin symmetry and $3D$ bosonization}}, \href{https://doi.org/10.1007/JHEP10(2020)007}{\emph{JHEP} {\bfseries 10} (2020) 007} [\href{https://arxiv.org/abs/1906.05834}{{\ttfamily 1906.05834}}].

\bibitem{Kalloor:2019xjb}
R.R.~Kalloor, \emph{{Four-point functions in large $N$ Chern-Simons fermionic theories}}, \href{https://doi.org/10.1007/JHEP10(2020)028}{\emph{JHEP} {\bfseries 10} (2020) 028} [\href{https://arxiv.org/abs/1910.14617}{{\ttfamily 1910.14617}}].

\bibitem{Silva:2021ece}
J.A.~Silva, \emph{{Four point functions in CFT\textquoteright{}s with slightly broken higher spin symmetry}}, \href{https://doi.org/10.1007/JHEP05(2021)097}{\emph{JHEP} {\bfseries 05} (2021) 097} [\href{https://arxiv.org/abs/2103.00275}{{\ttfamily 2103.00275}}].

\bibitem{Binder:2021cif}
D.J.~Binder, S.M.~Chester and M.~Jerdee, \emph{{ABJ Correlators with Weakly Broken Higher Spin Symmetry}}, \href{https://doi.org/10.1007/JHEP04(2021)242}{\emph{JHEP} {\bfseries 04} (2021) 242} [\href{https://arxiv.org/abs/2103.01969}{{\ttfamily 2103.01969}}].

\bibitem{Gerasimenko:2021sxj}
P.~Gerasimenko, A.~Sharapov and E.~Skvortsov, \emph{{Slightly broken higher spin symmetry: general structure of correlators}}, \href{https://doi.org/10.1007/JHEP01(2022)097}{\emph{JHEP} {\bfseries 01} (2022) 097} [\href{https://arxiv.org/abs/2108.05441}{{\ttfamily 2108.05441}}].

\bibitem{Jain:2022ajd}
P.~Jain, S.~Jain, B.~Sahoo, K.S.~Dhruva and A.~Zade, \emph{{Mapping Slightly Broken Higher Spin (SBHS) theory correlators to Free theory correlators: A momentum space bootstrap using SBHS symmetry}},  \href{https://arxiv.org/abs/2207.05101}{{\ttfamily 2207.05101}}.

\bibitem{Bzowski:2019kwd}
A.~Bzowski, P.~McFadden and K.~Skenderis, \emph{{Conformal $n$-point functions in momentum space}}, \href{https://doi.org/10.1103/PhysRevLett.124.131602}{\emph{Phys. Rev. Lett.} {\bfseries 124} (2020) 131602} [\href{https://arxiv.org/abs/1910.10162}{{\ttfamily 1910.10162}}].

\bibitem{Bzowski:2020kfw}
A.~Bzowski, P.~McFadden and K.~Skenderis, \emph{{Conformal correlators as simplex integrals in momentum space}}, \href{https://doi.org/10.1007/JHEP01(2021)192}{\emph{JHEP} {\bfseries 01} (2021) 192} [\href{https://arxiv.org/abs/2008.07543}{{\ttfamily 2008.07543}}].

\bibitem{Caloro:2022zuy}
F.~Caloro and P.~McFadden, \emph{{Shift operators from the simplex representation in momentum-space CFT}}, \href{https://doi.org/10.1007/JHEP03(2023)106}{\emph{JHEP} {\bfseries 03} (2023) 106} [\href{https://arxiv.org/abs/2212.03887}{{\ttfamily 2212.03887}}].

\end{thebibliography}\endgroup
\end{document}